\documentclass[11pt]{article}
\usepackage{amsmath,amssymb,hyperref,graphicx,color,xcolor,mathtools}
\usepackage{subcaption}
\usepackage[percent]{overpic}

\usepackage{cite}
\numberwithin{equation}{section}
\newcommand{\beq}{\begin{equation}}
\newcommand{\eeq}{\end{equation}}
\newcommand{\bea}{\begin{eqnarray}}
\newcommand{\eea}{\end{eqnarray}}
\newcommand{\beal}{\begin{equation}\begin{aligned}}
\newcommand{\eeal}{\end{aligned}\end{equation}}

\newcommand{\LL}{\mathcal{L}}
\newcommand{\HH}{\mathcal{H}}
\newcommand{\DD}{\mathcal{D}}
\newcommand{\OO}{\mathcal{O}}
\newcommand{\SSS}{\mathcal{S}}
\newcommand{\tr}{\mathrm{Tr}}

\newcommand{\mr}{\mathrm}

\begin{document}
\begin{titlepage}

\begin{center}

{\bf \Large Quasiparticle tunnelling in two coupled chiral SYK model}

\vspace{12mm}

\renewcommand\thefootnote{\mbox{$\fnsymbol{footnote}$}}
Avik Chakraborty${}^{1}$\footnote{avikchakraborty589@gmail.com},
Manavendra Mahato${}^{1}$\footnote{manav@iiti.ac.in}

\vspace{6mm}

\vspace{2mm}
${}^{1}${\small \sl Department of Physics} \\
{\small \sl Indian Institute of Technology Indore} \\
{\small \sl Khandwa Road, Indore - 453552, MP, India}

\end{center}

\vspace{12mm}

\noindent

 The chiral SYK model \cite{Lian_2019} is a $1+1$ dimensional generalisation of the Sachdev-Ye-Kitaev model with chiral Majorana fermions and homogeneous random interactions. In the large-$N$ limit, the model admits an exact solution of the two-point function due to its scaling symmetry and exhibits a quantised thermal Hall conductance consistent with that of a $2+1$-dimensional gapped topological system. We study two chiral SYK systems coupled by a relevant quadratic interaction that explicitly breaks scaling and time-reversal symmetry. Working in the regime of weak intersystem coupling, we solve the Dyson-Schwinger equations perturbatively and obtain analytic expressions for two-point functions at finite temperature. Unlike the coupled SYK model in $0+1$ dimensions, the $1+1$-dimensional chiral system does not develop a mass gap, and no thermal phase transition is observed. We show that the leading correction to the thermodynamic free energy is temperature independent, implying that the entropy density remains identical to that of two uncoupled chiral SYK systems. A real-time analysis of the retarded correlator reveals the emergence of massless collective bosonic modes propagating between the two subsystems at zero temperature, signalling quasiparticle tunnelling without gap generation. Our results demonstrate a sharp qualitative distinction between relevant deformations of SYK models in zero and one spatial dimensions, and highlight the robustness of gapless chiral edge dynamics against explicit scaling symmetry-breaking interactions.

\end{titlepage}
\setcounter{footnote}{0}
\renewcommand\thefootnote{\mbox{\arabic{footnote}}}

\hrule
\tableofcontents
\bigskip
\hrule

\addtolength{\parskip}{8pt}
%%%%%%%%%%%%%%%%%%%%%%%%%%%%%%%%%%%%%%%%
\section{Introduction}\label{sec:introduction}
%%%%%%%%%%%%%%%%%%%%%%%%%%%%%%%%%%%%%%%%
The Sachdev-Ye-Kitaev (SYK) model \cite{kitaevtalks,Sachdev_1992} has garnered significant attention in the study of holography due to its status as a solvable large-$N$ quantum field theory in one dimension (1D). As a prototypical example, it offers valuable insights into the nature of holography in lower-dimensional quantum mechanical settings. A notable characteristic of the SYK model is its acquisition of conformal symmetry in the low-energy limit, which is explicitly and spontaneously broken to nearly conformal quantum mechanics. This feature mirrors the behaviour of black holes approaching extremality, where a nearly AdS$_2$ region emerges \cite{Maldacena_2016,Kitaev_2017}. The SYK model demonstrates quantum chaotic behaviour, evidenced by the exponential growth of out-of-time-ordered four-point functions. It also serves as an excellent toy model of strange metals for its non-Fermi liquid behaviour \cite{Hartnoll_2018,Song_2017,Sachdev_2015}.

The SYK model is a quantum field theory (QFT) of $N$ Majorana fermions $\psi^i$ in $0+1$ dimensions with all-to-all $q$-fermion random interaction. The system is governed by the action
\beq
 S_{\mr{SYK}} = \int dt \left[ \frac{i}{2} \sum_{i} \psi^i \partial_t \psi^i - i^{q/2}\sum_{i_1  \ldots i_q} J_{i_1 \ldots i_q} \psi^{i_1} \ldots \psi^{i_q} \right],
\eeq
where $J_{i_1 \ldots i_q}$ are chosen randomly and independently from a Gaussian distribution with zero mean and $ \langle J_{i_1 \ldots i_q}^2 \rangle = (q-1)! J^2 /N^{q-1}$ in the large-$N$ limit. Hermiticity and time-reversal symmetry restrict $q/2$ to even numbers. The ground state of the system is highly degenerate in the large-$N$ limit leading to a large ground-state entropy which is also observed in gravitational settings like black hole horizon or in condensed matter system with spin glass like states. At low temperature, strong coupling, or in the IR limit, the system acquires a reparametrisation symmetry that is spontaneously broken to SL$(2,\mathbb{R})$ and the corresponding soft mode is governed by a Schwarzian action. The Schwarzian action also emerges as a boundary action of the AdS$_2$ throat of near extremal black holes \cite{Almheiri:2014cka,Jensen:2016pah,maldacena2016conf,Engels_y_2016}. The SYK model exhibits maximal chaos at low temperatures \cite{Gu_2019}. The out-of-time-ordered correlators (OTOCs) grow exponentially with time and the maximal Lyapunov exponent saturates the Maldacena-Shenker-Stanford bound \cite{Maldacena:2015waa}. 

There have been various proposals to generalise the SYK model into $1+1$ dimensions. These generalisations can be broadly classified into homogeneous and inhomogeneous models, depending on whether the SYK-type random couplings exhibit explicit dependence on the spatial coordinate. The introduction of an additional spatial dimension also allows for the emergence of fermionic chirality, enriching the possible physical phenomena. One class of $1+1$ D extensions involve SYK dots on $1$D lattices which have been extensively studied in the context of transport phenomena, quantum chaos and non-Fermi liquid behaviour \cite{Gu_2017,Berkooz_2017,PhysRevD.96.106008, PhysRevB.95.155131,Jian_2017,Chen_2017,Cai_2018,Zhang_2018}. Homogeneous nonchiral models involving interactions through random couplings or tensorial fields have also been studied in the large-$N$ limit \cite{Turiaci_2017,Murugan_2017,Berkooz_2017_thirring,Narayan_2017,Klebanov_2017,Giombi_2017}. In particular, a direct generalisation of the random coupling with nonchiral fermions is marginally irrelevant \cite{Berkooz_2017_thirring}. In this work, we specifically address \textit{the chiral SYK model}---an extension with chiral Majorana fermions interacting through homogeneous random 4-fermion interaction.

The chiral SYK model \cite{Lian_2019} consists of $N$ right chiral Majorana fermions $\psi^i$ in $1+1$ D with a marginal four-fermion all-to-all SYK interaction and is governed by the action
\beq
 \label{eq:action_chiralSYK}
 S_{\mr{chiral}}=\int dt dx \left(\frac{i}{2}\sum_{i}\psi^i(\partial_t+\partial_x)\psi^i +\sum_{i<j<k<l}J_{ijkl}\psi^i\psi^j\psi^k\psi^l\right)\ ,
\eeq
where the random Gaussian couplings $J_{ijkl}$ are spatially uniform. A canonical free fermion action assigns a scaling dimension of $1/2$ to the fermion fields. Consequently, a $q$-fermion interaction is marginal for $q=4$ and becomes irrelevant for $q>4$. At zero temperature, the presence of scaling symmetry allows for an exact solution of the Dyson-Schwinger (DS) equations. The random four-fermion interaction in the chiral SYK model remains exactly marginal, permitting an exact solution for the averaged two-point function across all energy scales. The four-fermion interaction explicitly breaks Lorentz invariance, resulting in spin-charge separation that persists at all energy scales. The two-point function factorises into a product of two $0+1$-dimensional SYK conformal propagators, each propagating with distinct velocities determined by the interaction strength, $u_\pm = 1 \pm J/2\pi$. At finite temperature, the exact large-$N$ correlator is obtained by exploiting the structural similarity between the zero-temperature solutions for $N=4$ and in the large-$N$ limit. These solutions remain valid for interaction strengths within the range $0 \le J < 2\pi$, beyond which UV nonlinearities become significant and cannot be neglected. The ground-state entropy, computed from the large-$N$ effective action using the exact solution for the propagators, coincides with that of $N$ free chiral Majorana fermions, indicating the absence of a large ground-state degeneracy in the model. An explicit evaluation of the energy-momentum tensor leads to a quantised thermal Hall conductivity that remains independent of the interaction strength. Furthermore, the four-point OTOCs exhibit exponential growth along spacetime trajectories with fixed velocity, with the Lyapunov exponent approaching the chaos bound as the interaction strength approaches its upper limit, $J \to 2\pi$. Although the maximal Lyapunov exponent saturates the chaos bound, the absence of large ground-state degeneracy contrasts with the expectations from holographic duals of black hole physics. Nevertheless, interacting chiral fermions of this type can exist as edge modes of $(2+1)$-dimensional gapped chiral topological phases. For example, chiral complex fermions, each equivalent to two chiral Majorana modes, appear at the edges of integer quantum Hall states \cite{laughlin1981quantized,halperin1982quantized} or Chern insulators \cite{haldane1988model,qi2006topological}. In contrast, an odd number of chiral Majorana fermions can arise at the boundaries of non-Abelian fractional quantum Hall states \cite{moore1991nonabelions,read2000paired}, $p+ip$ chiral superconductors \cite{qi2011topological}, or chiral spin liquids \cite{kitaev2006anyons}.

The chiral SYK model describes a QFT that allows for exact solutions of the zero temperature and thermal two-point functions across all energy scales. An important question is whether one can introduce a scale into the model while retaining its solvability. A relevant interaction Hamiltonian quadratic in fermionic fields explicitly breaks the scaling and time-reversal symmetry of the model. A natural choice is $\Delta \HH = 2i \sum_{i<j} A_{ij}(x)\psi^i \psi^j $, where $A_{ij}$ is antisymmetric and has spatial dependence in general. However, such an interaction term can be eliminated by the SO$(N)$ rotation $\psi^i \to (e^{-\int_{-\infty}^x A_{ij}(x')dx'}) \psi^j$. In this work, we construct a coupled system of two chiral SYK models interacting through a Hamiltonian
\beq
 \label{eq:int_Hamiltonian}
 \HH_{I} = i\mu \sum_i \psi^i_A (t,x) \psi^i_B (t,x),
\eeq
where $A$ and $B$ denote the two subsystems and $\mu$ is a dimensionful constant parametrising the interaction strength between the two systems. This setup is a direct generalisation of the $0+1$-dimensional coupled SYK model extensively studied in the context of traversable wormholes \cite{Maldacena_2017,maldacena2018eternal,maldacena2020syk,Lensky_2021}. The physics of the coupled SYK system in the IR limit is identical to the eternal traversable wormhole solution of nearly AdS$_2$. The quadratic interaction makes the wormhole traversable by setting up an interaction between the two boundaries that seems nonlocal in the bulk but may arise locally in higher-dimensional ambient space \cite{Gao_2017}. The coupled system shows a phase transition from the thermal AdS phase at low temperature to the black hole phase at high temperature. At low temperature, the system acquires a gap above the ground state which is evident from the exponential decay of two-point functions and the existence of quasiparticle peaks in the spectral function. The ground state of the coupled system is close to the thermofield double state of the two decoupled systems. Our goal is to introduce such quadratic interaction in $1+1$-dimensional chiral SYK system and analyse its consequences.

In this paper, we study two chiral SYK systems coupled by the relevant interaction Hamiltonian in Eq. \eqref{eq:int_Hamiltonian}. This Hamiltonian is a generic term for the breaking of time-reversal and scaling symmetry and serves as a representative of tunnelling interaction between the two subsystems $A$ and $B$. For the purpose of solvability, we restrict ourselves to the limit where the two chiral systems are weakly coupled. We obtain the perturbative solution of the two-point function analytically at finite temperature as a combination of complete elliptic integrals of the first kind by solving the DS equation using the exact two-point function of chiral SYK model. A consistent regularisation scheme is necessary to perform the real space integrals on the exact correlator. We show that the leading correction to the thermodynamic free energy is independent of the temperature; thus, the entropy density does not get any enhancement. Unlike the $0+1$-dimensional case, the correlators do not decay exponentially for large Euclidean time suggesting a gapless phase at all temperatures. For fixed spatial coordinate $x$, the averaged fermionic correlation between two subsystems follows the same profile as the averaged correlation between fermions of the same subsystem and the ratio of their strengths is independent of temperature. The retarded two-point function at finite temperature is computed from the Euclidean correlator by analytic continuation of the elliptic integrals. At zero temperature, the retarded correlator is logarithmic in nature, which is generally observed in the theory of massless free scalar fields in $1+1$ dimensions. An explicit computation of the spectral function at zero temperature validates the existence of massless bosonic excitations propagating between the two subsystems.

The chiral SYK model is effectively described by two massless free chiral bosons with modified ``speed of light" $u_\pm$ that depends explicitly on the SYK coupling strength $J$. The model admits a quantised thermal Hall conductivity independent of $J$ suggesting the topological nature of the chiral system. In a $2+1$-dimensional topological system, parity and time-reversal symmetry is broken in the presence of $p$-wave fermion pairing \cite{read2000paired,qi2011topological}. The quadratic interaction between two chiral SYK systems considered in this work is a generic deformation that softly breaks the scaling and time-reversal symmetry. Our results highlight a sharp qualitative distinction between coupled SYK models in $0+1$ and $1+1$ dimensions. While a relevant deformation in $0+1$-dimensional SYK model leads to a gapped phase and wormhole-like physics, the same deformation in a chiral edge theory preserves gaplessness and gives rise to propagating collective modes, which is expected in the theory of Majorana fermions living on the edge of $2+1$ dimensional topological bulk systems. This demonstrates that traversable-wormhole intuition does not straightforwardly extend to higher-dimensional chiral systems. We find evidence of massless collective bosonic modes, with velocities $u_\pm$ propagating between the two subsystems at zero temperature, which we identify as propagating quasiparticles between the two chiral subsystems. A real-time analysis is necessary for a stronger understanding of the implications of breaking time-reversal symmetry.

This paper is organised as follows. In Sec. \ref{sec:chiralsyk}, we briefly review the key ideas and results on the two-point function of chiral SYK model \cite{Lian_2019} for $N=4$ and large $N$. For $N=4$, the model can be solved via bosonisation. In the large-$N$ limit, the two-point functions can be obtained as the exact solution of Dyson-Schwinger equation. We discuss the implications of a consistent regularisation scheme in both the scenarios. Thermal properties of the chiral system have also been summarised. The readers familiar with this model can skip Sec. \ref{sec:chiralsyk} and proceed to Sec. \ref{sec:coupledchiralsyk} where we investigate two chiral SYK models coupled by a relevant bilinear coupling. In Sec. \ref{subsec:DS}, we obtain the DS equations in the large-$N$ limit. When the systems are weakly coupled, the leading correction to the two-point function can be obtained by solving the DS equation and is presented in Sec. \ref{subsec:Ncorrelators}. In Sec. \ref{subsec:thermal_quantities}, we calculate the leading correction to thermodynamic free energy using the two-point function. The leading correction to free energy is also computed using a direct approach verifying the solution for the two-point function computed in Sec. \ref{subsec:Ncorrelators}. The retarded correlator and the spectral function are computed in Sec. \ref{subsec:spectra}. Finally, we summarise the results and comment on a few future directions in Sec. \ref{sec:conclusion}.

%%%%%%%%%%%%%%%%%%%%%%%%%%%%%%%%%%%%%%%%
\section{Review of the chiral SYK model}\label{sec:chiralsyk}
%%%%%%%%%%%%%%%%%%%%%%%%%%%%%%%%%%%%%%%%
 In this section, we revisit the chiral SYK model and review the key ideas and results presented in Ref. \cite{Lian_2019}. It is a $1+1$-dimensional generalisation of the SYK model that involves $N$ flavours of right-chiral Majorana fermions, $\psi^i$, and is governed by the action
 \beq
  \label{eq:chiralSYK_action}
  S_{\mr{chiral}}=\int dt dx \left(\frac{i}{2}\sum_{i=1}^N\psi^i(\partial_t+\partial_x)\psi^i +\sum_{1\le i<j<k<l\le N}J_{ijkl}\psi^i\psi^j\psi^k\psi^l\right)\ ,
 \eeq
 where the fermions follow the anticommutation relation $\lbrace \psi^i(t,x),\psi^j(t,x') \rbrace = \delta^{ij}\delta(x-x')$ and the SYK couplings are randomly picked from a Gaussian distribution with $\langle J_{ijkl} \rangle = 0$ and
 \beq
  \langle J_{ijkl}J_{i'j'k'l'} \rangle = \frac{3! J^2}{(N-1)(N-2)(N-3)}\delta_{ii'}\delta_{jj'}\delta_{kk'}\delta_{ll'}\quad (i<j<k<l, \, i'<j'<k'<l').
 \eeq
 
 The fermion fields have a canonical scaling dimension 1/2. So, the four-fermion interaction is marginal, and the random couplings $J_{ijkl}$ are dimensionless. As has been shown in \cite{Lian_2019}, the interaction remains exactly marginal for all energy scales. Although the interaction breaks Lorentz symmetry, the scaling symmetry is intact. In Secs. \ref{subsec:N4prop} and \ref{subsec:largeNprop}, we shall see that the absence of Lorentz invariance in the model is not a threat to its solvability. The scaling symmetry plays a crucial role in computing the exact solutions for the two-point functions at zero temperature for both $N=4$ and large-$N$ cases.
 
 \subsection{point splitting regularisation and bosonisation for \texorpdfstring{$N=4$}{N=4}}\label{subsec:N4prop}
 
 To investigate any quantum field theoretic model, it is necessary to use a consistent regularisation scheme. The product of two quantum fields at the same spacetime point is regularised by computing the operator product expansion (OPE) at split spacetime points and taking the limit $(t',x')\rightarrow (t,x)$ keeping the product normal ordered. When Lorentz symmetry is absent, the regularisation depends on the direction of point splitting. Point splitting in the $x$ direction serves as a natural regularisation scheme in condensed matter theory and is employed throughout this paper. The product of two operators constructed from the fermions, $\OO_i$ and $\OO_j$, at coinciding spacetime points is regularised as
 \beq
  \OO^i(t,x)\OO^j(t,x) = \mathrm{lim}_{x' \rightarrow x} : \OO^i(t,x)\OO^j(t,x') :
 \eeq
 where $:\OO:$ stands for normal ordering of the operator $\OO$.
 
 The $1+1$ D chiral SYK model is integrable for a minimal number of flavors $N=4$ which can be observed via bosonisation. The model has only one coupling parameter $J_{1234}=J \geq 0$ for $N=4$ set as a constant. In terms of the chiral complex fermions defined as $c^\uparrow=(\psi^1+i\psi^2)/\sqrt{2}$ and $c^\downarrow=(\psi^3+i\psi^4)/ \sqrt{2}$, the action reads
 \beq
 \label{eq:cfermion_action}
  S_4 = \int dt dx \left(i\sum_{\sigma=\uparrow,\downarrow}{c^\sigma}^\dagger(\partial_t+\partial_x)c^\sigma -J n^\uparrow n^\downarrow\right)\ ,
 \eeq
 where $n^{\sigma}={c^{\sigma}}^{\dagger}c^{\sigma}$ is the fermion density for $\sigma = \uparrow, \downarrow$. It is necessary to impose normal ordering and a field redefinition $c^{\sigma}(t,x) \to e^{iJx/2}c^{\sigma}(t,x)$ in order to obtain the action $S_4$. Eq. \eqref{eq:cfermion_action} resembles the chiral Luttinger model with a spin-charge separation. The action is bosonised by expressing the complex fermions in terms of vertex operators in the manner
 \beq
  c^{\sigma}(t,x)=\eta^\sigma :e^{i\phi^\sigma(t,x)}:\ ,\qquad {c^{\sigma}}^\dagger(t,x)={\eta^\sigma}^\dagger :e^{-i\phi^\sigma(t,x)}:\quad (\sigma=\uparrow,\downarrow)\ ,
 \eeq
 where $\eta^{\sigma}$ are the Klein factors \cite{haldane1981luttinger,heidenreich1980luttinger,von_Delft_1998} necessary to recover the fermion anticommutation relation and $\phi^{\sigma}$ are two real scalar fields with $\phi = \phi + 2\pi$ identified. The bosons satisfy the equal-time commutation relation
 \beq
  [\phi^\sigma(t,x),\phi^{\sigma'}(t,x')]=i\pi\delta^{\sigma\sigma'}\text{sgn}(x-x')\ ,
 \eeq
 with $\text{sgn}(x-x')$ representing the sign of $x$. The normal-ordered products of the fermionic operators in the action \eqref{eq:cfermion_action} can be expressed in terms of the boson fields by regularising the OPEs imposing point splitting in the $x$ direction. Appendix A of Ref. \cite{Lian_2019} contains a detailed derivation of the identities,
 \beq
 : n^\sigma : \; = \; : {c^\sigma}^\dagger c^\sigma : \; = \; : \frac{\partial_x\phi^\sigma}{2\pi} : \ ,\qquad :-i{c^\sigma}^{\dagger} \partial_x c^\sigma : \; = \; : \frac{(\partial_x\phi^\sigma)^2}{4\pi} : \ .
 \eeq
 The bosonised action is obtained by substituting these identities in \eqref{eq:cfermion_action},
 \beq
 S_4=-\int dtdx\left(\frac{1}{4\pi}\sum_{\sigma=\uparrow,\downarrow}\partial_x\phi^\sigma(\partial_t\phi^\sigma +\partial_x\phi^\sigma)+ \frac{J}{4\pi^2}\partial_x\phi^\uparrow\partial_x\phi^\downarrow \right)\ .
 \eeq
 This describes a free theory of the bosons $\phi^{\sigma}$ since the bosons only appear as bilinears in the action. Furthermore, the bosons can be decoupled from each other by the field redefinition $\phi^{\pm} = (\phi^{\uparrow} \pm \phi^{\downarrow})/\sqrt{2}$. The action becomes
 \beq
 \label{eq:bosonised_action}
 S_4=-\int dtdx\left(\sum_{\alpha=\pm} \frac{1}{4\pi}\partial_x\phi^\alpha(\partial_t\phi^\alpha +u_\alpha\partial_x\phi^\alpha)\right).
 \eeq
 The action in Eq. \eqref{eq:bosonised_action} resembles the Floreanini-Jackiw action of two chiral scalar fields with velocities $u_\pm = 1 \pm J/2\pi$ implying a broken Lorentz symmetry. The SYK coupling coefficient $J$ modifies the ``speed of light" for the two scalars. The velocities $u_\pm$ are independent of the energy scale, and consequently the interaction is exactly marginal for $N=4$. Furthermore, positivity of the two velocities imply that the interaction strength should have a bound $ 0 \le J < 2\pi $. If $J > 2\pi$, the velocity $u_-$ becomes negative, flipping the chirality of $\phi^-$. If the chiral SYK model describes an edge theory of $2+1$-dimensional gapped bulk theory, the chirality must be protected topologically as long as the bulk gap does not close. To understand the theory for $J>2\pi $, irrelevant UV nonlinearities must be taken into account at high energies. In this paper, we restrict the analysis to interaction strength inside the physical bound $0 \le J < 2\pi $ and ignore any UV nonlinearities.
 
 \paragraph{Bosonic and fermionic two-point functions:}
 The bosonised theory for four flavours of Majorana is a free theory of two decoupled chiral bosons $\phi^{\pm}$. The two-point function of $\phi^{\pm}$ (without time ordering) can be directly computed from the action \eqref{eq:bosonised_action}. At zero temperature,
 \beq
 \label{eq:bosonic_correlator}
 \langle \phi^\alpha(t,x)\phi^{\alpha'}(0,0)\rangle =-\delta^{\alpha\alpha'}\log\left[2\pi i(u_\alpha t-x-i0^+)\right],
 \eeq
 where $\alpha,\alpha'=\pm$ and $0^+$ denotes a positive infinitesimal real number. The two-point function at a finite temperature $\beta^{-1}$ can be obtained by summing over the zero-temperature two-point functions at time $t+im\beta$ ($m \in \mathbb{Z}$),
 \beq
 \label{eq:bosonic_correlator_beta}
 \langle \phi^\alpha(t,x)\phi^{\alpha'}(0,0)\rangle_\beta =-\delta^{\alpha\alpha'}\log\left\{2i\beta u_\alpha \sinh\left[\frac{\pi}{\beta}\left(t-u_\alpha^{-1}x-i0^+\right)\right]\right\}
 \eeq
 The bosonic two-point functions are sufficient to compute all higher-point functions in the bosonised theory using Wick's theorem. In addition, the fermionic correlators can be expressed in terms of the bosonic correlators. The average two-point function of the Majorana fermions (without time ordering) for $N=4$  at temperature $\beta^{-1}$ is defined as
 \beq
  G_\beta(t,x)=\frac{1}{4}\sum_{i=1}^4\langle \psi^i(t,x)\psi^i(0,0)\rangle_\beta .
 \eeq
 The Majorana fermions are expressed in terms of the complex fermions, $c^{\sigma}$ and ${c^\sigma}^\dagger$, which in turn can be replaced by the vertex operators $:e^{i\phi^\sigma}:$ and $:e^{-i\phi^\sigma}:$, respectively. Then by applying Wick's theorem, the averaged fermion correlator is rewritten as
 \beq
 \label{eq:exp_bosonic_corr}
 G_\beta(t,x) = \frac{1}{2}\sum_{\sigma= \uparrow,\downarrow}e^{\langle\phi^\sigma(t,x)\phi^\sigma(0,0)\rangle_\beta} = \prod_{\alpha=\pm} e^{\frac{1}{2}\langle \phi^{\alpha}(t,x)\phi^{\alpha}(0,0) \rangle_{\beta} }.
 \eeq
 Now, it is straightforward to compute the fermionic correlators at zero and finite temperature by substituting the bosonic correlators \eqref{eq:bosonic_correlator} and \eqref{eq:bosonic_correlator_beta} into Eq. \eqref{eq:exp_bosonic_corr},
 \bea
 \label{eq:ferm_correlator}
  G(t,x) &=& \frac{1}{2\pi i}\left[ (u_+ t-x-i0^+)(u_- t-x-i0^+) \right]^{-1/2}, \\
  \label{eq:ferm_correlator_temp}
  G_\beta(t,x) &=& \frac{1}{2i\beta\sqrt{u_+u_-}}\left\{\sinh\left[ \frac{\pi}{\beta}(t-u_+^{-1}x-i0^+)\right] \sinh\left[ \frac{\pi}{\beta}(t-u_-^{-1}x-i0^+)\right]\right\}^{-1/2}.
 \eea
  It is easy to check that in the limit $\beta \rightarrow \infty$, Eq. \eqref{eq:ferm_correlator_temp} reduces to Eq. \eqref{eq:ferm_correlator}.
 
 \paragraph{Thermal quantities:} Various thermodynamic quantities can be obtained by computing the energy-momentum tensor $ T^\mu_\nu$ from the bosonic (Eq. \eqref{eq:bosonised_action}) as well as the fermionic action (Eq. \eqref{eq:chiralSYK_action}) for $N=4$. For the fermionic action, it is necessary to impose the point splitting in the $x$ direction ($t \rightarrow 0,x \rightarrow \epsilon$ with $\epsilon \rightarrow 0$) in order to regularise operator products at coinciding spacetime point, whereas the choice of point splitting direction does not affect the physical quantities for the bosonised action. The infinitesimal parameter $\epsilon$ helps separate the diverging vacuum expectation values. Substracting the diverging pieces, we find the physical energy density and energy current as
 \beq
  \mathcal{E}= \langle T_0^0 \rangle_{\beta} =\frac{\pi}{12\beta^2}(u_+^{-1}+u_-^{-1})\ ,\qquad j_\mathcal{E}= \langle T_0^x \rangle_{\beta} = \frac{\pi}{6\beta^2}\ .
 \eeq
 The energy current gives a thermal Hall conductance independent of $J$, $\kappa_{xy} = \partial j_\mathcal{E} / \partial (\beta^{-1}) = \pi /3\beta$, which is in agreement with the general belief that thermal Hall conductance is topologically invariant and quantised as $\kappa_{xy} = N\pi /12 \beta$ for $N=4$ chiral Majorana fermions living on the edge of gapped $2+1$ D topological system \cite{kane1997quantized}. Since the zero-temperature entropy density of free bosons is zero, the thermal entropy density $\SSS_4$ can be derived from the energy density using $\beta^{-1}[\partial \SSS_4 /\partial(\beta^{-1})]=\partial \mathcal{E}/\partial(\beta^{-1})$, which gives $\SSS_4 =(\pi/6\beta)(u_+^{-1}+u_-^{-1})$.

 \subsection{Large \texorpdfstring{$N\ $}\ two-point functions and UV regularisation}\label{subsec:largeNprop}
  In this subsection, the chiral SYK model is analysed in the large-$N$ limit using the Dyson-Schwinger equation. For convenience, let us introduce the Euclidean time coordinate, $\tau = it$, to study the model at finite temperature. The effective theory at large $N$ is described by the two-point Green's function averaged over the Majorana flavour indices. At zero temperature, it is defined as
  \beq
   G(\tau,\tau';x,x') \equiv \frac{1}{N}\sum_{i=1}^N\langle T \psi^i(\tau,x)\psi^i(\tau',x')\rangle ,
  \eeq
  where $T$ stands for imaginary time ordering. For finite temperature $\beta^{-1}$, the definition becomes
  \beq
  G^{\beta}(\tau,\tau';x,x') = \frac{1}{N}\sum_{i=1}^NZ^{-1}\text{Tr}\left[e^{-\beta H}T\psi^i(\tau,x)\psi^i(\tau',x')\right] ,
  \eeq
  where $Z=\text{Tr} e^{-\beta H}$ is the partition function of the system. Because of the translational symmetry of the model, the two-point function is only dependent on the difference between spacetime coordinates of the two-points, $G(\tau,\tau';x,x') = G(\tau-\tau';x-x')$. Without losing generality $\tau' = 0 = x'$ can be chosen throughout the analysis. Since the fermions are anticommuting, the averaged Green's function is antisymmetric $G(-\tau,-x) = -G(\tau,x)$. Also, the cyclic property of trace renders $G$ antiperiodic $G(\tau + \beta,x) = -G(\tau,x)$ with a period of the inverse temperature $\beta$.
  
  When the four-fermion interaction is turned off, the free theory is described by $N$ decoupled right-chiral Majorana fermions in $1+1$ dimensions. Thus the imaginary time-ordered averaged two-point function of the free fermions at zero temperature is derived to be
  \beq
   G^f(\tau,x)=\frac{1}{N}\sum_{i=1}^N\langle T \psi^i(\tau,x)\psi^i(0,0)\rangle_f=\frac{1}{2\pi} \frac{1}{\tau-ix},
  \eeq
  with its fourier transform
  \beq
   G^f(\omega_\tau,k)=\int d\tau dx e^{-i\omega_\tau\tau-ikx} G^f(\tau,x)=\frac{-i}{\omega_\tau-ik}.
  \eeq
  Similar to the one-dimensional SYK model, only the melonic diagrams contribute to the two-point function $G$ to the leading order in $1/N$. The DS equation for $G$ and the self-energy $\Sigma$ effectively describes the theory in the large-$N$ limit,
  \beq
   \label{eq:chiral_DSeqn}
   \frac{1}{G(\omega_\tau,k)}=\frac{1}{G^f(\omega_\tau,k)} -\Sigma(\omega_\tau,k)\ ,\qquad\qquad \Sigma(\tau,x)=J^2[G(\tau,x)]^3 ,
  \eeq
   where the first equation is in the momentum space ($\omega_\tau,k$) of ($\tau,x$). Since $(G^f)^{-1}$ has the same scaling dimension as $\Sigma$ at zero temperature, one cannot ignore the free propagator term at low energy unlike the $0+1$-dimensional SYK model. Consequently, the DS equation does not possess a conformal symmetry in the IR.
   
   \paragraph{Zero temperature solution:}
    At zero temperature, the bilinears $G$ and $\Sigma$ have scaling dimensions $-1$ and $+1$, respectively. Because of the exact scaling invariance, $G$ and $\Sigma$ can be constrained down to the following ansatz based on their scaling dimensions,
  \beq
  \label{eq:ansatz}
  \Sigma(\omega_\tau,k)=\kappa f(e^{i\theta_k})\ ,\qquad \qquad G(\omega_\tau,k)=\frac{-i}{\kappa} \frac{1}{e^{-i\theta_k}+if(e^{i\theta_k})}\ ,
  \eeq
  where $(\kappa,\theta_k)$ are the polar coordinates in momentum space, $\omega_\tau +ik = \kappa e^{i\theta_k} $, and as $\Sigma$ is antisymmetric, $f(z) = -f(-z)$ is an odd function defined on the unit circle $|z|=1$. It is possible to find an exact form of $f(z)$ at zero temperature by finding the inverse Fourier transform of the ansatz $G(\omega_\tau,k)$ and $\Sigma(\omega_\tau,k)$ in terms of $f$ and its derivatives, then plugging them into the second DS equation. A detailed and thorough analysis is presented in Ref. \cite{Lian_2019}. We summarise only the elements essential to our work. Subject to the assumption that $|zf(z)|<1$ and $f(z)$ has no intrinsic singularity for $|z|<1$, $f(z)$ follows a second-order ordinary differential equation
  \beq
   \label{eq:diff_eqn_f}
   f(z)-zf'(z)-z^2f''(z)=\frac{J^2}{4\pi^2}\frac{-iz^3}{[1+izf(z)]^3},
  \eeq
  where $f'$ and $f''$ are the first and second derivatives of $f$, respectively. Eq. \eqref{eq:diff_eqn_f} has a general solution,
  \beq
   \label{eq:soln_f_gen}
   f(z)=iz^{-1}\left [1-\sqrt{{\frac{1}{c_0}}-{\frac{c_0 J^2}{16\pi ^2}}(z^2-s_0)^2}\right ]
  \eeq
  with two undetermined constants $c_0$ and $s_0$. When the system is free ($J = 0$), the self-energy vanishes, so $f(z)$ should also vanish identically. This fixes $c_0 = 1$. The other constant $s_0$ depends on the choice of point splitting. For a point splitting in the $x$ direction, the two-point function $G$ should reduce to the free two-point function $G(0,x) = G^f(0,x) = \frac{i}{2\pi x}$ on a constant time slice $\tau = 0$. This imposes a constraint $f(z)=0$ at $z=\pm 1$ fixing $s_0 = 1$. The solution to $f(z)$ for an $x$ directional point splitting becomes
  \beq
   \label{eq:soln_f}
   f(z)=iz^{-1}\left[1-\sqrt{1-\frac{J^2}{16\pi^2}(z^2-1)^2}\right] .
  \eeq
  The assumption on $f(z)$ dictates that $J$ must satisfy a bound $0 \le J <2\pi $ for the validity of the solution. An exact form of $f(z)$ implies an exact solution of the DS equation. The two-point function and self-energy is obtained using \eqref{eq:soln_f} in position space. In the large-$N$ limit,
   \bea
   \label{eq:chiral_corrG}
   G(\tau,x) &=& \frac{1}{2\pi}\frac{1}{\sqrt{(u_+\tau-ix)(u_-\tau-ix)}}, \\
   \Sigma(\tau,x) &=& \frac{J^2}{8\pi^3}\frac{1}{\left[(u_+\tau-ix)(u_-\tau-ix)\right]^{3/2}},
  \eea
  where $u_\pm = 1 \pm J/2\pi$ are the two velocities. Because of the square roots in $G$ and $\Sigma$, the branch should be specified to uniquely determine the functions. A choice consistent with the point splitting in $x$ direction is $G(\tau,x)\rightarrow \frac{i}{2\pi x}$ and $\Sigma(\tau,x)\rightarrow -\frac{iJ^2}{8\pi^3 x^3}$ when $|x| \gg |\tau|$ with the branch cut lying on a straight line segment from $-iu_+ \tau$ to $-iu_- \tau$. This uniquely determines the functions everywhere except the singular point $\tau=0=x$.
  
  The two-point functions for $N=4$ [Eq. \eqref{eq:ferm_correlator}] and large $N$ (Eq. \eqref{eq:chiral_corrG}) look exactly the same, except $J$ has different meaning in the two cases. For $N=4$, $J$ is the only coupling constant in the model, whereas for large $N$, $J$ is proportional to the square root of the Gaussian average amplitude $\langle J^2_{ijkl}\rangle$.

 \paragraph{Real-space UV regularisation:} The choice of point splitting plays a crucial role in determining the two-point Green's functions in both $N=4$ and large-$N$ scenarios. While performing integrals over the real space, $|\tau| \geq \epsilon$ is the correct UV cutoff consistent with the point splitting in the $x$ direction. To justify that, consider the Fourier transformations of $G$ and $\Sigma$ with a real-space UV cutoff $|\tau| \geq \epsilon$,
 \beq
  W_\sigma(\omega_\tau,k)=\int_{-\infty}^\infty d\tau e^{-i\omega_\tau\tau}\Theta(|\tau|-\epsilon) \int_{-\infty}^\infty dx e^{-ikx} (u_+\tau-ix)^\sigma(u_-\tau-ix)^\sigma\ ,
 \eeq
 where $\Theta$ is the Heaviside theta function and $\sigma = -1/2,-3/2$ recovers momentum space $G(\omega_\tau,k)=\frac{1}{2\pi}W_{-\frac{1}{2}}(\omega_\tau,k)$ and $\Sigma(\omega_\tau,k)=\frac{J^2}{8\pi^3}W_{-\frac{3}{2}}(\omega_\tau,k)$, respectively. Without losing generality, we can consider $k>0$. The $x$ integral is then equivalent to a contour integral on the real $x$ axis closing in the lower half-plane. The integrand has branch points at $x=-iu_\pm \tau$, so the branch cut lies inside the contour for $\tau > 0$; otherwise, the contour integral is zero. Deforming the contour around the branch cut from $-iu_+ \tau$ to $-iu_- \tau$ and defining $y=ix$, the Fourier transform is expressed as 
 \beq
  W_\sigma(\omega_\tau,k)=-2\sin(\pi\sigma)\int_{\epsilon}^\infty d\tau e^{-i\omega_\tau\tau} \int_{u_-\tau}^{u_+\tau} dy e^{-ky} (u_+\tau-y)^\sigma(y-u_-\tau)^\sigma.
 \eeq
 The resulting double integral can be decoupled by a change of variables, $y_+=\frac{u_+\tau-y}{u_+-u_-}$ and $y_-=\frac{y-u_-\tau}{u_+-u_-}$, and is expressed as a combination of complete and incomplete Gamma functions. Finally, the Fourier transform can be expanded in a power series of the cutoff parameter $\epsilon$ and regularised by taking $\epsilon \rightarrow 0$. It turns out that the Fourier transform of the two-point function does not require a UV regularisation and is given by
 \beq
  \label{eq:fourierG}
  G(\omega_\tau,k)=\frac{1}{2\pi}W_{-\frac{1}{2}}(\omega_\tau,k) =\frac{-i}{\sqrt{(\omega_\tau-iu_+k)(\omega_\tau-iu_-k)}}.
 \eeq
 For $\sigma = -3/2$, the UV cutoff gives rise to a regular piece $i\omega_\tau +k$. The Fourier transform of the self-energy can be expressed as
 \beq
  \label{eq:fourierSigma}
  \Sigma(\omega_\tau,k)=\frac{J^2}{8\pi^3}W_{-\frac{3}{2}}(\omega_\tau,k)
=i\omega_\tau+k -i\sqrt{(\omega_\tau-iu_+k)(\omega_\tau-iu_-k)}
 \eeq 
 It is easy to check that the Fourier transforms, Eqs. \eqref{eq:fourierG} and \eqref{eq:fourierSigma}, satisfy the DS equation Eq. \eqref{eq:chiral_DSeqn}.
 
 A different regularisation scheme $|x|>\epsilon$ for the real space integrals corresponds to $s_0 =-1$ in the solution \eqref{eq:soln_f_gen}. If we start from a solution of $G$ and $\Sigma$ with $s_0=-1$ and follow the regularisation procedure with a point splitting in $\tau$, we shall find the same Fourier transforms as Eq. \eqref{eq:fourierG} and Eq. \eqref{eq:fourierSigma} that satisfy the DS equation. The solutions of the bilinears in real space depend on the choice of regularisation scheme.

\begin{figure}[ht]
   \centering
   \includegraphics[scale=0.55]{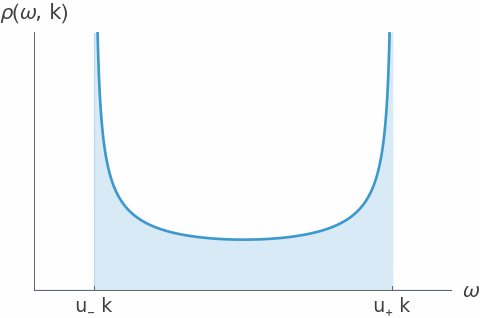}
   \caption{The spectral density of states of the chiral SYK model for fixed momentum $k >0$.}
   \label{spectral_chiral}
  \end{figure}

 Similar to the $N=4$ case, the two-point function and its Fourier transform are only valid in the regime $0 \le J<2\pi$, beyond which the total chirality will not be preserved unless UV nonlinearities are taken into consideration. It can also be observed from the spectral density of states which can be obtained from the retarded correlator as
 \beq
  \label{eq:spectral_dos_chiral}
  \rho(\omega,k) = 2 \; \mr{Im}\left[ G_{\mr{ret}}(\omega +i0^+,k) \right] = \frac{2 \; \theta(u_+ k -\omega) \theta(\omega -u_- k)}{\sqrt{(u_+ k -\omega)(\omega -u_- k)}},
 \eeq
 where $\theta$ represents the Heaviside theta function. For fixed momentum $k$, the energy $E_n$ of the many-particle eigenstate $\left| n,k \right\rangle$ is distributed in the range $E_n \in [u_- k, u_+ k]$ as shown in Fig. \ref{spectral_chiral}. For $J>2\pi$, one of the velocities becomes negative, $u_-<0$, which leads to eigenstates with reversed chirality. To avoid considering the UV nonlinearities, the SYK coupling must be restricted to $0 \le J<2\pi$.
 
 \paragraph{Finite-temperature solution:}
  We now proceed to discuss the two-point function of chiral SYK model at a finite temperature $\beta^{-1}$. At a nonzero temperature, the Euclidean time $\tau$ becomes periodic in $\beta$ and the frequencies $\omega_\tau$ are replaced by discreet Matsubara frequencies $\omega_n$. Unlike the zero-temperature case, $\beta^{-1}$ introduces a energy scale to the system and the scaling invariance of DS equation is lost; thus, the ansatz Eq. \eqref{eq:ansatz} is no longer applicable. However, there are hints from the zero-temperature solution that helps to determine the finite-temperature correlator. First, the averaged two-point function in the large-$N$ limit (Eq. \eqref{eq:chiral_corrG}) has exactly the same form as the averaged fermionic correlator for $N=4$ (Eq. \eqref{eq:ferm_correlator}). The second hint is that, at zero temperature, $G$ and $\Sigma$ factorises into the conformal two-point function and self-energy solutions of two $0+1$-dimensional SYK model, respectively, along $\tau - iu_+^{-1}x$ and $\tau - iu_-^{-1}x$ directions. Although such a factorisation is invalid in the UV, the discussion on momentum space $G$ and $\Sigma$ suggests that it is plausible in the IR. Motivated from these observations, it can be guessed that the finite-temperature two-point function in the large-$N$ limit has the same form as Eq. \eqref{eq:ferm_correlator_temp} for $N=4$,
  \beq
   \label{eq:finitecorr}
G^{\beta}(\tau,x) = \frac{1}{2\beta \sqrt{u_+ u_-}} \frac{1}{\sqrt{ \sin \left[ \frac{\pi}{\beta}(\tau-iu_+^{-1}x) \right] \sin \left[ \frac{\pi}{\beta}(\tau-iu_-^{-1}x) \right] } }.
  \eeq
  
  As a verification of Eq. \eqref{eq:finitecorr}, it must be shown that the Fourier transforms of $G^{\beta}$ and $\Sigma^{\beta} = J^2 (G^{\beta})^3$ satisfy the second DS equation in Eq. \eqref{eq:chiral_DSeqn}. For convenience, the coordinates are redefined by setting $\beta = 2\pi$. The results are restored by rescaling $(\omega_n,k)\rightarrow \frac{\beta}{2\pi} (\omega_n,k)$ for a generic temperature $\beta^{-1}$. Introducing the proper UV cutoff $|\tau| \geq \epsilon$, a computation of the Fourier transformation
  \beq
  \label{eq:finite_fourier}
   W_\sigma(\omega_n,k)=\int_{\epsilon}^{2\pi-\epsilon} \frac{d\tau}{2\pi} e^{-i\omega_n\tau} \int_{-\infty}^\infty dx e^{-ikx} \left[\sin\left(\frac{\tau-iu_+^{-1}x}{2}\right) \sin \left(\frac{\tau-iu_-^{-1}x}{2}\right)\right]^\sigma
  \eeq
  leads to the two-point function $G^\beta(\omega_{n},k)=\frac{1} {2\sqrt{u_+u_-}}W_{-1/ 2}(\omega_{n},k)$ and self-energy $\Sigma^\beta(\omega_n,k)=\frac{J^2}{32\pi^2 (u_+u_-)^{3/2}}W_{-3/2}(\omega_{n},k)$ in the momentum space.
  
  Unlike the zero-temperature case, the integrand here has infinitely many branch cuts in the complex $x$ plane for a given $\tau$. Thus, the integral must be treated carefully to obtain the Fourier transforms of the bilinears. This is done by transforming the imaginary time integral to real-time integrals and looking at the cancellations of branch cut contributions. A detailed analysis is presented in Ref. \cite{Lian_2019}. The integral is regular for $\sigma=-1/2$ and leads to the momentum space two-point function at finite temperature $\beta^{-1}$,
  \beq
   G^\beta(\omega_n,k)=\frac{i\beta \sqrt{u_+u_-}}{2J} \frac{\Gamma\left[\frac{\beta u_+u_-}{2J}(-\omega_n u_+^{-1}+ik)+\frac{1}{4}\right]\Gamma\left[\frac{\beta u_+u_-}{2J}(-\omega_n u_-^{-1}+ik)+\frac{1}{4}\right]}{ \Gamma\left[\frac{\beta u_+u_-}{2J}(-\omega_n u_+^{-1}+ik)+\frac{3}{4}\right] \Gamma\left[\frac{\beta u_+u_-}{2J}(-\omega_n u_-^{-1}+ik)+\frac{3}{4}\right]}\ ,
  \eeq
  For $\sigma = -3/2$, the UV cutoff $\epsilon$ produces a regular piece $i\omega_n +k$. The momentum space self-energy is obtained as
  \beq
   \Sigma^\beta(\omega_n,k)=i\omega_n+k+\frac{2iJ}{\beta \sqrt{u_+u_-}} \frac{\Gamma\left[\frac{\beta u_+u_-}{2J}(-\omega_n u_+^{-1}+ik)+\frac{3}{4}\right]\Gamma\left[\frac{\beta u_+u_-}{2J}(-\omega_n u_-^{-1}+ik)+\frac{3}{4}\right]}{ \Gamma\left[\frac{\beta u_+u_-}{2J}(-\omega_n u_+^{-1}+ik)+\frac{1}{4}\right] \Gamma\left[\frac{\beta u_+u_-}{2J}(-\omega_n u_-^{-1}+ik)+\frac{1}{4}\right]}.
  \eeq
  These momentum space expressions of the two-point function $G^\beta(\omega_n,k)$ and self-energy $\Sigma^\beta(\omega_n,k)$ satisfy the DS equation in Eq. \eqref{eq:chiral_DSeqn}.
 
 \paragraph{Thermal quantities:}
 Thermodyamic quantities for the chiral SYK system in the large-$N$ limit can be derived from the finite-temperature two-point function in Eq. \eqref{eq:finitecorr} following a procedure similar to the procedure for the $N=4$ case. The energy density and energy current are obtained as
 \beq
  \mathcal{E} = \langle T^0_0 \rangle_{\beta} = \frac{N\pi}{48\beta^2}(u_+^{-1}+u_-^{-1}) , \qquad j_\mathcal{E}= \langle T^x_0 \rangle_{\beta} = \frac{N\pi}{24\beta^2} .
 \eeq
 This gives a thermal Hall conductance of $N\pi /12 \beta$ when the system is an edge of a $2+1$-dimensional gapped topological state. This is in agreement with the expectation that thermal Hall conductance is independent of $J$ and quantised at order $\OO(N)$. The entropy density is derived from $\beta^{-1}[\partial \mathcal{S}/\partial(\beta^{-1})]=\partial \mathcal{E}/\partial(\beta^{-1})$, giving
 \beq
  \label{eq:entropy1}
  \SSS = \SSS_0 + \left[ \frac{N\pi}{24} (u_+^{-1} + u_-^{-1}) \right] \beta^{-1},
 \eeq
 where $\SSS_0$ is the ground-state entropy density. The entropy density of the system can also be found from the disorder-averaged partition function,
 \beq
  \langle Z_{\mr{chiral}} \rangle = \int \prod_{1\le i<j<k<l\le N}dJ_{ijkl}e^{- N^3J^2_{ijkl}/(2\cdot3!J^2)}\int\mathcal{D}\psi^i e^{-S_\text{chiral}[\psi^i(\tau,x),J_{ijkl}]} \; .
 \eeq
 After taking the Gaussian average over $J_{ijkl}$, the two-point function $\bar{G}$ and the self-energy $\bar{\Sigma}$ are introduced in the path integral by inserting the functional identity 
 \beq
 \int \DD \bar{G} \DD \bar{\Sigma} \exp\left[ -\frac{N}{2} \int d\tau dx \int d\tau' dx' \bar{\Sigma}\left(\bar{G}-\frac{1}{N}\sum_i \psi^i(\tau,x)\psi^i(\tau',x') \right) \right] = 1 .
 \eeq
 Integrating over the fermionic fields, the partition function is expressed as a functional integral of the two-point functions $\bar{G}$ and $\bar{\Sigma}$. The thermodynamic free energy and entropy density for system size $L$ can be obtained to leading order in $1/N$ by substituting the on-shell values, $G$ and $\Sigma$, in the expressions
  \beq
   -\frac{\beta F_{\mr{chiral}}}{L} = \frac{\log \langle Z_{\mr{chiral}} \rangle}{L} \quad \mr{and} \quad \SSS_{\mr{chiral}} = (1-\beta \partial_{\beta}) \frac{\log \langle Z_{\mr{chiral}} \rangle}{L},
  \eeq
  where  
  \beq
  \label{eq:chiral_logZ}
   \frac{\log \langle Z_\mr{chiral} \rangle}{L} = \frac{N}{2L} \tr \log \left[ (\partial_{\tau}-i\partial_x) -\Sigma \right] - \frac{N}{2L}\int d\tau dx d\tau' dx' \left( \Sigma G - \frac{J^2}{4} G^4 \right).
  \eeq
  Note that the physically relevant quantity is $\langle \log Z_{\mr{chiral}} \rangle$. However the difference between $\langle \log Z_{\mr{chiral}} \rangle $ and $ \log \langle Z_{\mr{chiral}} \rangle $ is of $\mathcal{O}(N^{-2}) $ \cite{Kitaev_2017}. To avoid dealing with the $\tr \log$ term, it is convenient to compute $\partial_{J}\log \langle Z_{\mr{chiral}} \rangle$ and integrate over $J$ to obtain the difference between the free energies of free chiral Majoranas and chiral SYK Majoranas. Since the coefficients of $\partial_J G$ and $\partial_J \Sigma$ vanish at the saddle point, we find
  \beal
  \label{eq:G4integral}
   \frac{\partial_{J} \log \langle Z_\mr{chiral} \rangle}{L} &= \frac{N\beta J}{4} \int_{-\beta/2}^{\beta/2} d\tau \int_{-\infty}^{\infty} dx \, {G(\tau,x)}^4 \\
   &= \frac{NJ}{64\beta^3 u_+^2 u_-^2} \int_{-\beta/2}^{\beta/2} d\tau \int_{-\infty}^{\infty} dx \dfrac{1}{\sin^2 \left[ \frac{\pi}{\beta}(\tau-iu_+^{-1}x) \right]\sin^2 \left[ \frac{\pi}{\beta}(\tau-iu_-^{-1}x) \right]}.
  \eeal
  A real space UV cutoff $(x/ \epsilon_x)^2 + (\tau / \epsilon_\tau)^2 \geq 1$ with $\epsilon_\tau / \epsilon_x \rightarrow 0$ is necessary to make the above integral convergent. Since the above integrand is an even function in $\beta$, the expression in Eq. \eqref{eq:G4integral} is an odd function in $\beta$. Therefore,
  \beq
   \label{eq:JJ_int}
   \int_0^{J} dJ \frac{\partial_J\log Z}{L} = a_0(\epsilon_{\tau},\epsilon_x,J)\beta + a_1(J) \beta^{-1} + a_2(\epsilon_{\tau},\epsilon_x,J)\beta^{-3} + \ldots,
  \eeq
  where the constants $a_j$ are of order $\mathcal{O}(\epsilon_{\tau,x}^{\, 2j-2})$. The diverging piece $a_0$ contributes to the zero-temperature free energy of the system. $a_2$ and higher pieces are vanishing in the limit $\epsilon_{\tau,x} \rightarrow 0$. Only $a_1$ is finite and does not depend on the cutoff parameters. $\partial_{J}a_1$ can be computed directly from \eqref{eq:G4integral} using a cutoff $| \tau | \geq \epsilon$ instead of the more complicated cutoff mentioned before (Appendix \ref{App:chiral_free_energy}), which gives $\partial_{J}a_1  = \frac{N}{96}(u_+^{-2} - u_-^{-2})$. The entropy density of $N$ free ($J=0$) chiral Majorana fermions is given by $\SSS_f = \SSS_{f0} + N\pi /12\beta$, where $\SSS_{f0}$ is the ground-state entropy density. This leads to an expression of the entropy density in the large-$N$ chiral SYK system,
  \beq
   \label{eq:entropy2}
   \SSS_{\mr{chiral}} = \SSS_{f0} + \left[ \frac{N\pi}{24} (u_+^{-1} + u_-^{-1}) \right] \beta^{-1} + \OO(\epsilon_x^2 \beta^{-3}).
  \eeq
  Eq. \eqref{eq:entropy2} agrees with Eq. \eqref{eq:entropy1}, verifying the results. The ground-state entropy density $\SSS_{f0}$ of $N$ free chiral Majorana fermions depends on the spatial boundary condition. For the antiperiodic boundary condition, the ground state is unique which leads to $\SSS_{f0} = 0$. For the periodic boundary condition, the ground state has an order $2^N$ degeneracy, thus the entropy density $\SSS_{f0} = \frac{N}{L}\log 2$. When the system size $L$ is parametrically larger than $N$, the ground-state entropy density of the chiral system vanishes for both cases, $\SSS_0 = \lim_{\beta \to \infty} \SSS_{\text{chiral}} = \SSS_{f0} = 0$.

%%%%%%%%%%%%%%%%%%%%%%%%%%%%%%%%%%%%%%%%
\section{Two coupled chiral SYK model}\label{sec:coupledchiralsyk}
%%%%%%%%%%%%%%%%%%%%%%%%%%%%%%%%%%%%%%%%

 \subsection{Model}
 In this article we investigate two chiral SYK systems, $A$ and $B$, coupled by an quadratic interaction Hamiltonian of the form
 \beq
  \HH_{\mr{I}} = i\mu \sum_{j=1}^{N} \psi^j_A \psi^j_B, \quad \HH = \HH_A + \HH_B + \HH_{\mr{I}},
 \eeq
 where $\HH_A $ and $ \HH_B$ are the Hamiltonian densities of the chiral system $A$ and system $B$ respectively. This is a direct generalisation of the coupled SYK \cite{maldacena2018eternal} to $1+1$ dimensions with right-moving chiral Majorana fermions, $\psi^j_A$ and $\psi^j_B$, which obey the anticommutation relation $\lbrace \psi^i_A(t,x),\psi^j_B(t,x') \rbrace = \delta^{ij}\delta(x-x')$. We will restrict ourselves to the case where $N$ is large and the two chiral SYK models have identical couplings, $J_{ijkl}^A = J_{ijkl}^B = J_{ijkl}$, randomly picked from a Gaussian distribution with
 \beq
  \langle J_{ijkl} \rangle = 0, \quad \langle J_{ijkl}^2 \rangle = \frac{3! J^2}{N^3}.
 \eeq
 The coupled model has the Lagrangian density
  \beq
  \label{eq:coupled_lagrangian}
   \LL = \frac{i}{2} \sum_{a=A,B} \ \sum_{j=1}^{N}  \psi^j_a(\partial_t+\partial_x)\psi^j_a  + \sum_{a=A,B} \ \sum_{1 \le i<j<k<l\le N}  J_{ijkl}^a \psi^i_a\psi^j_a\psi^k_a\psi^l_a 
-i\mu \sum_{j=1}^{N} \psi_A^j \psi_B^j.
  \eeq
 The bilinear coupling breaks the scaling symmetry of chiral SYK model and introduces a scale to the model. Furthermore, the interaction does not respect the time-reversal symmetry of the chiral SYK system. The Euler-Lagrange equations of motion are easily found to be
 \bea
  \label{eq:eom1}
  i \left( \partial_t + \partial_x \right) \psi_A^i + \sum_{j<k<l} J_{ijkl} \psi^j_A\psi^k_A\psi^l_A -i\mu \psi_B^i &=& 0 \; , \\
  \label{eq:eom2}
  i \left( \partial_t + \partial_x \right) \psi_B^i + \sum_{j<k<l} J_{ijkl} \psi^j_B\psi^k_B\psi^l_B +i\mu \psi_A^i &=& 0 \; .
 \eea
 The Hamiltonian density is obtained from the Legendre transformation of the Lagrangian density as
 \beq
  \label{eq:hamiltonian}
  \HH = -\frac{i}{2} \sum_{a=A,B} \ \sum_{j=1}^{N}  \psi^j_a \partial_x \psi^j_a  - \sum_{a=A,B} \ \sum_{1 \le i<j<k<l\le N}  J_{ijkl}^a \psi^i_a\psi^j_a\psi^k_a\psi^l_a  +i\mu \sum_{j=1}^{N} \psi_A^j \psi_B^j.
 \eeq
 In terms of the complex fermions fields $c_a^j = (\psi_a^{2j} + i\psi_a^{2j+1})/\sqrt{2}$, the quadratic term represents a tunnelling interaction between the two subsystems. In the bosonised picture, $c_a^j = :e^{i\phi_a^j}:$, we can represent the interaction as
 \beq
  \HH_I = i\mu \sum_{j=1}^{N/2} \left( {c_A^j}^\dagger {c_B^j} + {c_A^j} {c_B^j}^\dagger \right)  = i\mu \sum_{j=1}^{N/2} \left( :e^{i\phi_A^j}::e^{-i\phi_B^j}: + :e^{-i\phi_A^j}::e^{i\phi_B^j}: \right).
 \eeq
 In this work, we shall restrict ourselves to small values of $\mu$ such that the tunnelling interaction remains weak compared to the random four-fermion interaction. This assumption will allow us to compute the averaged correlators analytically at leading order in $\mu$.
 
 \subsection{Large-\texorpdfstring{$N\ $}\ equations}\label{subsec:DS}
  In this subsection, we compute the effective equations governing the dynamics of the coupled model in the large-$N$ limit. The disorder-averaged partition function can be written as
  \beq
   \langle Z \rangle = \int \DD J_{ijkl}^a \DD \psi_a^j \ e^{i\int dt dx \LL} \ \ \mr{with} \ \ \DD J_{ijkl}^a = \prod_{a=A,B}\ \prod_{1 \le i<j<k<l\le N} dJ_{ijkl}^a \exp \left( -\frac{J^2_{ijkl}}{2\frac{3!J^2}{N^3}}\right).
  \eeq
  The Gaussian average over the random interactions is obtained by integrating over $J_{ijkl}^a$. We perform the Hubbard-Stratonovich transformation to reduce the fermionic system to a theory of the collective fields $G$ and $\Sigma$. The dynamical mean field, $\bar{G}_{ab}(\tau,x;\tau',x') = \frac{1}{N}\sum_{j=1}^{N}\psi^j_a(\tau,x) \psi^j_b(\tau',x')$, is introduced in the path integral by inserting the identity
  \beal
   1 &= \int \mathcal{D}\bar{G}_{ab} \delta \left( N\bar{G}_{ab} - \sum_{j=1}^{N}\psi^j_a\psi^j_{b} \right) \\
&= \int \mathcal{D}\bar{G}_{ab} \mathcal{D}\bar{\Sigma}_{ab} \exp \left[-\frac{1}{2}\int d\tau dx d\tau' dx' \ \bar{\Sigma}_{ab}\left( N\bar{G}_{ab} - \sum_{j=1}^{N}\psi^j_a\psi^j_{b} \right) \right],
  \eeal
  where the self-energy $\Sigma_{ab}$ acts as the Lagrange multiplier enforcing the delta constraint. The averaged partition function can be rewritten as a functional of the bilinears, $\bar{G}_{ab}$ and $\bar{\Sigma}_{ab}$, by integrating out the Majorana fermion fields $\psi^j_a$,
  \beq
   \label{eq:partition_function}
   \langle Z \rangle _J=\int \mathcal{D}\bar{G}_{ab} \mathcal{D}\bar{\Sigma}_{ab} \ e^{-NI_{\mr{eff}}(\bar{G}_{ab},\bar{\Sigma}_{ab})},
  \eeq
  where the effective action,
  \beal
  \label{eq:effective_action}
   I_{\mathrm{eff}}= -\frac{1}{2} \log \det \left[\delta_{ab}(\partial_{\tau}-i\partial_x) -\bar{\Sigma}_{ab}\right] + \frac{1}{2} \int &d\tau dx d\tau' dx' \sum_{a,b=A,B}\Big(\bar{\Sigma}_{ab}\bar{G}_{ab}- \frac{J^2}{4} \bar{G}_{ab}^4\Big)\\ &+\frac{i\mu}{2} \int d\tau dx \Big( \bar{G}_{AB}-\bar{G}_{BA}\Big).
  \eeal
  In the limit $N \rightarrow \infty$, the effective action \eqref{eq:effective_action} describes the coupled system. The DS equations are obtained from the effective action by varying $I_{\mathrm{eff}}$ with respect to the bilinears $\bar{G}_{ab}$ and $\bar{\Sigma}_{ab}$,
  \bea
   \label{eq:dseqn1}
   G_{a b} &=& {G_0}_{a b}+\left({G_0} \star \Sigma \star G\right)_{ab},\\
   \label{eq:dseqn2}
   \Sigma_{ab} &=& J^2 \left(G_{ab}\right)^3 -i\mu_{ab} \delta(\tau - \tau')\delta(x-x'),
  \eea
  where $G$ and $\Sigma$ denote the on-shell values of the bilinears, $G_0$ denotes the free propagator with $\left( G_0^{-1} \right)_{ab} = \delta_{ab}(\partial_{\tau}-i\partial_x)$, $\mu_{ab}=\begin{pmatrix}
	0 & \mu \\ -\mu & 0
  \end{pmatrix}$ and $\star$ represents a convolution,
  \beq
   \left(\Sigma \star G \right)_{ab}\left(\tau,x;\tau',x'\right) = \int d\tau_1 dx_1 \sum_{c=A,B} \Sigma_{ac}\left(\tau,x;\tau_1,x_1\right)G_{cb}\left(\tau_1,x_1;\tau',x'\right).
  \eeq
   The DS equations, Eqs. \eqref{eq:dseqn1} and \eqref{eq:dseqn2}, can be restricted further by imposing the symmetries of the system. Since the model has translational symmetry, the averaged two-point functions depend on the difference between the time and space coordinate, $G_{ab}(\tau,x;\tau',x') = G_{ab}(\tau-\tau',x-x')$. Without losing generality, we can choose $\tau' = 0$ and $x'=0$. Consequently, it is clear from \eqref{eq:dseqn2} that the self-energies $\Sigma_{ab}$ also possess the translational invariance. Systems $A$ and $B$ are taken to be identical; thus, $G_{AA}=G_{BB}$. The two-point function follows the antisymmetry condition $G_{ab}(\tau,x) = -G_{ba}(-\tau,-x)$ which can be realised from the anticommutation relation of the fermions. Finally, the set of DS equations reduces to
  \bea
  \label{eq:DS1}
   G_{AA} &=& {G_0}_{AA} + (G_0 \star \Sigma \star G)_{AA}, \\
   \label{eq:DS2}
   G_{AB} &=& (G_0 \star \Sigma \star G)_{AB}, \\
   \label{eq:DS3}
   \Sigma_{AA} &=& J^2 G_{AA}^3, \\
   \label{eq:DS4}
   \Sigma_{AB} &=& J^2 G_{AB}^3 - i\mu \delta(\tau)\delta(x).
  \eea
  
  At finite temperature $\beta^{-1}$, the two-point function is antiperiodic with period $\beta$. Furthermore, there is an exact $\mathbb{Z}_4$ symmetry to the Lagrangian \eqref{eq:coupled_lagrangian}. This together with the antiperiodicity constrains $G_{AB}$ to be antisymmetric about $\tau=\beta/2$. We will not restrict ourselves by imposing this condition on $G_{AB}$, but in Sec. \ref{subsec:Ncorrelators}, we shall see that the solution of $G_{AB}$ in the limit $\mu^2 \ll J$ respects this constraint.
  
 \subsection{Two-point function at finite temperature}\label{subsec:Ncorrelators}
  
  Here we evaluate $G_{AB}$ at finite temperature for small $\mu$. For $\mu = 0$, the two copies of chiral SYK are decoupled and the correlators are the conventional chiral SYK correlator \eqref{eq:finitecorr} at finite temperature $1/\beta$,
  \beq
   \label{eq:propzeromu}
   G_{AA}^{\beta}=G^{\beta}=G_{BB}^{\beta}, \quad G_{AB}^{\beta}=0=G_{BA}^{\beta}.
  \eeq
  where $G^{\beta}$ is the finite-temperature two-point function of the chiral SYK model given in Eq. \eqref{eq:finitecorr}. Now, if we turn on a small value of $\mu$, the correlation between the $A$ and $B$ system is no longer zero. The leading correction to both the correlators $G_{AA}^{\beta}$ and $G_{BB}^{\beta}$ is of order $\mu^2$. Restricting ourselves to the linear order in $\mu$, only $G_{AB}^\beta$ and $G_{BA}^\beta$ receive corrections from the DS equations \eqref{eq:DS2} and \eqref{eq:DS4},
  \beq
   \label{eq:GLRcorrection}
   G_{AB}^{\beta}(\tau,x) = i\mu \int_{0}^{\beta}d\tau_1 \int_{-\infty}^{\infty}dx_1\  G_{AA}^{\beta}(\tau-\tau_1,x-x_1)G_{BB}^{\beta}(\tau_1,x_1),
  \eeq
  The real space integral needs a consistent regularisation scheme as discussed in Sec. \ref{subsec:largeNprop}. We present the explicit analytic computation of the convolution in Appendix \ref{App:two-point_fn} and directly write the result here. The leading correction to $G_{AB}^{\beta}(\tau,x)$ is expressed as
 \beq
  \label{eq:GAB_finitetemp}
  G_{AB}^{\beta}(\tau,x) = \frac{2\mu}{\pi^2 (u_+ -u_-)}  \left[ \sqrt{\frac{c_+}{c_-}} K\left( \frac{1}{c_-} \right)K\left( c_+ \right) - \sqrt{\frac{c_-}{c_+}} K\left( \frac{1}{c_+} \right)K\left(c_- \right)\right],
 \eeq
 where 
 \beq
  c_\pm = \exp \left[ \frac{\pi}{\beta}(u_\pm^{-1}x +i\tau) \right]
 \eeq
 and $K(c)$ denotes complete elliptic integral of the first kind as defined in Eq. \eqref{eq:elliptic_integral}.
 We verify the analytic expression of $G_{AB}^\beta$ in Sec. \ref{subsec:thermal_quantities} by computing the leading correction to thermodynamic free energy. Here, we focus on gaining more intuition about the solution of the two-point function. Notice that all the time dependence of the two-point function lies inside the elliptic integrals $K$. The UV divergence of the correlator at coincident spacetime points can be observed from the branch point singularities of $K(c)$ at $c= \pm 1$. $K$ is a multivalued function in the complex $c$ plane. When $c$ is real and $|c|<1$, Eq. \eqref{eq:elliptic_integral} defines the principle value of $K(c)$. The function can be analytically continued as a single-valued function in the whole complex $c$ plane except the singular branch point $c=1$ by making cuts from $c=1$ to $c \to \infty$ and from $c=-1$ to $c \to -\infty$ \cite{NIST:DLMF,lawden2013elliptic}.

  \begin{figure}[ht]
  \centering
  \begin{subfigure}[b]{0.45\textwidth}
    \centering
    \includegraphics[scale=0.52]{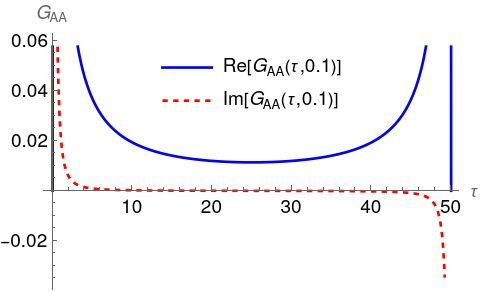}
  \end{subfigure}
  \hspace{0.3cm}
  \begin{subfigure}[b]{0.45\textwidth}
    \centering 
    \includegraphics[scale=0.52]{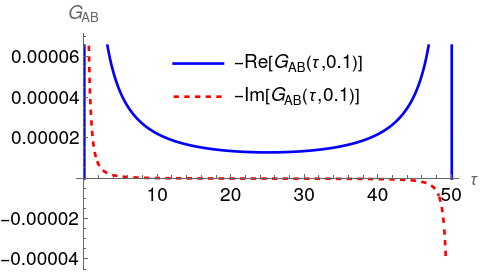}
  \end{subfigure}
  \caption{ The real and imaginary parts of the two-point functions are shown for parameter values $\beta = 50$, $J = 3$ and $\mu = 0.01$. For fixed $x$, the correlators follow same profile with different strengths.}
  \label{propagator_plots}
\end{figure}

  The solutions for the two-point functions, $G_{AA}^\beta = G^\beta$ from Eq.\eqref{eq:finitecorr} and $G_{AB}^\beta$ from Eq.\eqref{eq:GAB_finitetemp}, are valid in the regime $0 \leq J < 2\pi$ and $\mu^2 \ll J$. Although the correlators $G_{AA}^\beta$ and $G_{AB}^\beta$ differ in magnitude, they exhibit identical behavior for fixed $x$, as shown in Fig.~\ref{propagator_plots}. This feature persists at all temperatures $\beta^{-1}$. The two-point functions do not decay exponentially at large Euclidean time, indicating that the spectrum remains gapless even at low temperatures. However, for fixed $\tau$, the correlators behave differently for small $x$, but both decay at large $x$.

  \paragraph{Consequence of $\mathbb{Z}_4$ symmetry:} There is a $\mathbb{Z}_4$ symmetry present in the system which can be realised at the Lagrangian level. It is easy to see that the Lagrangian in Eq. \eqref{eq:coupled_lagrangian} is invariant under the discrete transformations $(\psi_A,\psi_B) \rightarrow (-\psi_B,\psi_A),(-\psi_A,-\psi_B) \; \mr{and} \; (\psi_B,-\psi_A)$. The symmetry puts constraints on the two-point function. In addition to antisymmetry and antiperiodicity, the correlator $G^{\beta}_{AA}$ and $G^{\beta}_{BB}$ are symmetric whereas $G^{\beta}_{AB}$ and $G^{\beta}_{BA}$ are antisymmetric around $\tau=\beta /2$,
  \beq
  \label{eq:Z4_symmetry_conditions}
   G^{\beta}_{AA}(\beta - \tau , -x) = G^{\beta}_{AA}(\tau,x) \quad \mr{and} \quad G^{\beta}_{AB}(\beta - \tau , -x) = -G^{\beta}_{AB}(\tau,x).
  \eeq
  At the leading order in $\mu$, $G^{\beta}_{AA} = G^{\beta}_{\mr{chiral}}$. It is straightforward to check from Eq. \eqref{eq:finitecorr} that $G^{\beta}_{\mr{chiral}}$ is symmetric around $\tau = \beta /2$. For $G^{\beta}_{AB}$, notice that $c_\pm (\beta-\tau,-x) = -c^{-1}_\pm (\tau,x)$. Substituting this into the expression of $G^{\beta}_{AB}$ in Eq. \eqref{eq:GAB_finitetemp}, we find that the antisymmetry condition in Eq. \eqref{eq:Z4_symmetry_conditions} is satisfied. In this subsection, we have not assumed that the $\mathbb{Z}_4$ symmetry is respected throughout our analysis. Still we find that the symmetry remains unbroken in the effective theory of bilinears, at least to leading order in $\mu$.
 
 \subsection{Thermal quantities}\label{subsec:thermal_quantities}
  
  In this subsection, we compute the leading correction to thermodynamic free energy density $F/L$ from the path integral using saddle-point approximation. Within replica-symmetric ansatz, we work with the annealed disorder average and compute $-\beta F/L = \log \langle Z \rangle /L$, where $L$ is the system size and $\langle Z \rangle$ is the disorder-averaged partition function expressed in Eq. \eqref{eq:partition_function}. Substituting the on-shell values of the two-point functions, we can write
  \beal
  \label{eq:coupled_logZ}
   \frac{\log \langle Z \rangle}{L} = \frac{N}{2L} \tr \log \left[ \delta_{ab}(\partial_{\tau}-i\partial_x) -\Sigma_{ab} \right] &- \frac{N}{2L}\int d\tau dx d\tau' dx' \Bigg[\sum_{a,b}\left( \Sigma_{ab}G_{ab}- \frac{J^2}{4} G_{ab}^4 \right) \\
   &+ i\mu \delta(\tau-\tau')\delta(x-x') \left(G_{AB} - G_{BA} \right)\Bigg].
  \eeal
  In the limit $\mu^2 \ll J$, the first two terms in the right-hand side of Eq.  \eqref{eq:coupled_logZ} reduces to twice of $\log \langle Z_{\mr{chiral}} \rangle$ in Eq. \eqref{eq:chiral_logZ} and the free energy density of the coupled system takes the form
  \beq
   \frac{F}{L} = \frac{2F_{\mr{chiral}}}{L} + \frac{\Delta F}{L}, \quad \mr{with} \quad \frac{\Delta F}{L} = \frac{i\mu N}{2} \left[ G_{AB}(0,0)-G_{BA}(0,0) \right].
  \eeq
  Here, we elaborate two techniques to determine the regularised value of the correlators: the first contains point splitting regularisation of the expression of $G_{AB}$ in Eq. \eqref{eq:GAB_finitetemp}, and the second involves solving the integral in Eq. \eqref{eq:GLRcorrection} with $\tau=0=x$. A consistent point splitting regularisation will yield the same result from both techniques. It also serves as a check to the analytic result we obtained in Eq. \eqref{eq:GAB_finitetemp}.
  
  As mentioned in Sec. \ref{subsec:Ncorrelators}, $K(c)$ is uniquely defined in the whole complex $c$ plane except the singular branch points $c=\pm 1$ by introducing a cuts from $c=1$ to $c \to \infty$ and $c=-1$ to $c \to -\infty$ \cite{NIST:DLMF,lawden2013elliptic}. When $c \to 1^-$, the function shows logarithmic divergence,
  \beq
   \lim_{c \to 1^-} \left[ K(c) - \frac{1}{2}\log \left( \frac{16}{1-c^2}\right) \right] = 0.
  \eeq
  When $c \to 1^+$, the value of $K(c)$ depends on whether we approach the branch cut from above or below. When Re$(c)>0$, $K(c)$ is analytically continued to $|c|>1$ using the Legendre's relation \cite{NIST:DLMF}
  \beq
  \label{eq:legendre_relation}
   K\left(\frac{1}{c}\right) = c\left[ K(c) \mp iK(c') \right],
  \eeq
  where the upper sign applies if Im$(c^2)>0$ and the lower sign applies if Im$(c^2)<0$ with the complementary modulus defined as $c'=\sqrt{1-c^2}$. Imposing a UV cutoff on the two-point function $G^{\beta}_{AB}(\epsilon_\tau,\epsilon_x)$ with $\epsilon_\tau,\epsilon_x >0$, determines the elliptic integrals in Eq. \eqref{eq:GAB_finitetemp} approaching the branch point singularity. Since $|c_\pm(\epsilon_\tau,\epsilon_x)|>1$ and Im$[c^2_\pm(\epsilon_\tau,\epsilon_x)]>0$, we can write
  \bea
    \label{eq:expansion_ell_int1}
    K\left[c_\pm^{-1}(\epsilon_\tau,\epsilon_x)\right] &=& \frac{1}{2}\log \left( \frac{8 \beta u_\pm}{\pi \epsilon_x} \right) + \mathcal{O}(\epsilon_x), \\
    \label{eq:expansion_ell_int2}
    K\left[c_\pm^{2}(\epsilon_\tau,\epsilon_x)\right] &=& -\frac{i\pi}{2} + \frac{1}{2}\log \left( \frac{8\beta u_\pm}{\pi\epsilon_x} \right) + \mathcal{O}(\epsilon_x).
  \eea
  where we have used $\epsilon_\tau/\epsilon_x \to 0$ consistent with the point splitting in $x$ direction. For $K\left[c_\pm^{2}(\epsilon_\tau,\epsilon_x)\right]$, the branch cut is approached from the above. The value differs from $K\left[c_\pm^{-2}(\epsilon_\tau,\epsilon_x)\right]$ by an amount $-iK(0)=-i\pi/2$. Substituting Eqs. \eqref{eq:expansion_ell_int1} and \eqref{eq:expansion_ell_int2} into Eq. \eqref{eq:GAB_finitetemp}, the divergent pieces are eliminated, and we arrive at a regularised value of the two-point function approaching the UV divergent point,
  \beq
  \label{eq:G_AB_regular}
   \left. G_{AB}^{\beta}(0,\epsilon_x) \right|_{\epsilon_x \rightarrow 0^+} = \frac{i\mu }{2\pi (u_+ -u_-)} \log \left( \frac{u_+}{u_-} \right).
  \eeq
  Since the Green's function is antisymmetric, we can substitute $G_{BA}^{\beta}(0,0)=-G_{AB}^{\beta}(0,0)$. Hence, the leading correction to the free energy density is obtained as
  \beq
   \label{eq:free_ener_corr1}
   \frac{\Delta F}{L} = -\frac{\mu^2 N}{2 J} \log \left( \frac{u_+}{u_-} \right).
  \eeq
  
  We can verify the above expression by computing $G_{AB}(0,0)$ directly from Eq. \eqref{eq:GLRcorrection},
  \beal
  \label{eq:G_AB_zero}
   G_{AB}^{\beta}(0,0) &= -i\mu \int_{\epsilon}^{\beta-\epsilon}d\tau \int_{-\infty}^{\infty}dx\  G_{AA}^{\beta}(\tau,x)G_{BB}^{\beta}(\tau,x) \\
   &= -\frac{i\mu}{4\beta^2 u_+u_-} \int_{\epsilon_\tau}^{\beta -\epsilon_\tau}d\tau \int_{-\infty}^{\infty} \frac{dx}{\sin \left[ \frac{\pi}{\beta}(\tau-iu_+^{-1}x) \right] \sin \left[ \frac{\pi}{\beta}(\tau-iu_-^{-1}x) \right]},
  \eeal
  where we have imposed a real space UV cutoff $|\tau| \geq \epsilon_\tau$ on the imaginary-time integral consistent with the point splitting in the $x$ direction. The explicit calculation of Eq. \eqref{eq:G_AB_zero} is presented in Appendix \ref{App:free_en_correction}. The leading-order correction to the free energy is obtained as
  \beq
  \label{eq:free_ener_corr2}
   \frac{\Delta F}{L} = -\frac{\mu^2 N}{2 J} \log \left( \frac{u_+}{u_-} \right).
  \eeq
  The expression in Eq. \eqref{eq:free_ener_corr2} completely agrees with Eq. \eqref{eq:free_ener_corr1}. The leading correction to the free energy is independent of temperature. Consequently, there is no contribution to the entropy density at leading order in $\mu^2/J$. It remains same as the entropy density of $2N$ free chiral Majorana fermions.

  The energy density $T^0_0$ is obtained from the Legendre transformation of the Lagrangian. The thermal expectation value can be obtained using $\HH$ in Eq. \eqref{eq:hamiltonian}. Imposing the equations of motion \eqref{eq:eom1} and \eqref{eq:eom2}, it is reduced to
  \beq
   \langle T^0_0 \rangle_\beta = \langle \HH \rangle_\beta = \frac{1}{4} \left\langle i \sum_{i,a}\psi_a^i (\partial_t -\partial_x) \psi_a^i + 2i\mu \sum_{i} \psi_A^i \psi_B^i \right\rangle_\beta .
  \eeq
  With the two-point functions $G_{AA}^\beta$ and $G_{AB}^\beta$ we can obtain the energy density at temperature $\beta^{-1}$ by performing a point splitting in the $x$ direction,
  \beq
   \langle T^0_0 \rangle_\beta = -\frac{iN}{4}(\partial_t -\partial_x)\left. (G_{AA}^\beta+G_{BB}^\beta)\right|_{t=0,x=\epsilon} + \frac{i\mu N}{2} \left. G_{AB}^\beta\right|_{t=0,x=\epsilon}.
  \eeq
  The first term represents the energy density of two decoupled chiral SYK systems. The enhancement due to $\mu$ comes from the second term involving $G_{AB}^\beta$. Using Eq. \eqref{eq:G_AB_regular}, the energy density is obtained as
  \beq
    \langle T^0_0 \rangle_\beta = \frac{N\pi}{24\beta^2}(u_+^{-1}+u_-^{-1}) - \frac{\mu^2 N}{4 J} \log \left( \frac{u_+}{u_-} \right).
  \eeq
  Since $u_+ >u_-$, the leading order correction to free energy $F$ and energy density $\langle T^0_0 \rangle_\beta$ suggests that the ground-state energy is lowered in the presence of the bilinear interaction.

\subsection{Retarded correlator and spectral density of states}\label{subsec:spectra}

In this subsection, we compute the retarded two-point function from the Euclidean time-ordered correlator $G_{AB}^{\beta}$ by analytic continuation,
\beal
 G_{AB,\mr{ret}}^{\beta}(t,x) &= \; \frac{\theta(t)}{N} \sum_{j=1}^{N} \langle \lbrace \psi^j_A(t,x),\psi^j_B(0,0) \rbrace \rangle \\
 &= \; \theta(t) \lim_{\epsilon \to 0} \left[ G_{AB}^{\beta}(it+\epsilon,x) - G_{AB}^{\beta}(it-\epsilon,x) \right],
\eeal
where $\theta(t)$ is the Heaviside theta function. We define the coordinates $x_\pm = \frac{\pi}{\beta}(t-u_\pm^{-1}x)$ and express the elliptic integrals $K(c)$ as a single valued function using the relation \eqref{eq:legendre_relation},
\beq
\label{eq:analytic_cont}
 \lim_{\epsilon \to 0} K\left( e^{X \pm i\epsilon} \right) = \theta(-X) K\left( e^{X} \right) + \theta(X) e^{-X} \left[ K\left( e^{-X} \right) \pm i K\left( \sqrt{1-e^{-2X}} \right) \right].
\eeq
Notice that, the presence of the heaviside theta functions ensure that the elliptic integrals contribute to the expression when the modulus is less than unity. Using $X=\pm x_{\pm}$ in Eq. \eqref{eq:analytic_cont} we obtain the retarded correlator at finite temperature as
\beal
 G_{AB,\mr{ret}}^{\beta}&(t,x) = \frac{4i\mu}{\pi J}\; \theta(t) \times \\
  \times &\Big[\theta(x_+)\theta(x_-)e^{-\frac{1}{2}(x_+ +x_-)} \lbrace K\left(e^{-x_-} \right) K\left(\sqrt{1-e^{-2x_+}} \right) - K\left(e^{-x_+} \right) K\left(\sqrt{1-e^{-2x_-}} \right) \rbrace \\
 +& \theta(x_+)\theta(-x_-)e^{-\frac{1}{2}(x_+ -x_-)}\lbrace K\left(e^{x_-} \right) K\left(\sqrt{1-e^{-2x_+}} \right) - K\left(e^{-x_+} \right) K\left(\sqrt{1-e^{2x_-}} \right) \rbrace \\
 +& \theta(-x_+)\theta(x_-)e^{\frac{1}{2}(x_+ -x_-)}\lbrace K\left(e^{-x_-} \right) K\left(\sqrt{1-e^{2x_+}} \right) - K\left(e^{x_+} \right) K\left(\sqrt{1-e^{-2x_-}} \right) \rbrace \\
 +& \theta(-x_+)\theta(-x_-)e^{\frac{1}{2}(x_+ +x_-)}\lbrace K\left(e^{x_-} \right) K\left(\sqrt{1-e^{2x_+}} \right) - K\left(e^{x_+} \right) K\left(\sqrt{1-e^{2x_-}} \right) \rbrace \Big].
\eeal
The correlator can be simplified further by introducing the variables $\tilde{x}_{\pm} = |x_{\pm}| = \frac{\pi}{\beta}|t-u_{\pm} x|$ and can be expressed as
\beq
 G_{AB,\mr{ret}}^{\beta}(t,x) = \frac{4 i \mu}{\pi J}\; \theta(t) e^{-\frac{1}{2}(\tilde{x}_+ +\tilde{x}_-)} \left[ K\left(e^{-\tilde{x}_-} \right) K\left(\sqrt{1-e^{-2\tilde{x}_+}} \right) - K\left(e^{-\tilde{x}_+} \right) K\left(\sqrt{1-e^{-2\tilde{x}_-}} \right) \right].
\eeq
It is straightforward to take the $\beta \to \infty$ limit in $G_{AB,\mr{ret}}^{\beta}(t,x)$ to obtain the retarded correlator at zero temperature, which yields
\beq
\label{eq:retarded_zero}
 G_{AB,\mr{ret}}(t,x) = \frac{i\mu}{J}\; \theta(t) \left[ \log|t-u_+^{-1}x -i0^+| - \log|t-u_-^{-1}x -i0^+| \right],
\eeq
where we have used the $i0^+$ prescription consistent with the retarded two-point function. The logarithmic correlator suggests that at zero temperature the effective modes propagating between the two subsystems are bosonic and massless. In momentum space,
\beq
 \tilde{G}_{AB,\mr{ret}}(\omega,k) = \int_{-\infty}^{\infty} dt \; e^{i\omega t -\eta t} \int_{-\infty}^{\infty} dx \; e^{ikx} G_{AB,\mr{ret}}(t,x),
\eeq
where $\eta \to 0^+$ is necessary to make the integral convergent when $t \to \infty$. The integrand is analytic in the upper half complex $x$ plane. For $k>0$, we can argue that the integral over $x$ is zero by closing the contour with a semicircle in the upper half-plane. However, for $k<0$, we must close the contour in the lower half-plane which yields a nonzero contribution due to the branch cut from $x=u_-t -i0^+$ to $x=u_+t -i0^+$. The retarded correlator in Fourier space becomes
\beq
\label{eq:retarded_fourier}
 \tilde{G}_{AB,\mr{ret}}(\omega + i0^+,k) = \frac{2\pi \mu}{J} \; \frac{\theta(-k)}{k} \left( \frac{1}{\omega +u_+k +i0^+} - \frac{1}{\omega +u_-k +i0^+} \right).
\eeq
The correlator has poles at $\omega = -u_\pm k -i0^+$ and infrared divergence when $k \to 0$. The inverse Fourier transform of the expression in Eq. \eqref{eq:retarded_fourier} involves exponential integrals. A consistent IR cutoff leads to the regularised value of the retarded correlator in Eq. \eqref{eq:retarded_zero}. The spectral function $\rho_{AB}$ can be extracted from the imaginary part of the momentum space retarded correlator. At zero temperature,
\beq
 \rho_{AB}(\omega,k) = 2 \; \mr{Im}\left[ \tilde{G}_{AB,\mr{ret}}(\omega + i0^+,k) \right] = \frac{4 \pi^2 \mu}{J} \;\frac{\theta(-k)}{k} \left[ \delta(\omega +u_-k) - \delta(\omega +u_+k) \right].
\eeq
The delta functions signify massless excitations at $\omega = -u_\pm k$ for $k<0$. The $\theta(-k)$ factor reflects the right-moving chirality of the Majorana fermions. The spectral weight $\rho_{AB}$ is fundamentally different from $\rho_{AA}$ obtained in Eq. \eqref{eq:spectral_dos_chiral}, which indicates the existence of many-body eigenstates within each subsystem. Unlike the Schwarzian-dominated dynamics of coupled SYK quantum mechanics, the interaction between the two chiral subsystems here is governed by massless collective bosonic modes with linear dispersion. This reflects the fundamentally different role of chirality in one spatial dimension.

%%%%%%%%%%%%%%%%%%%%%%%%%%%%%%%%%%%%%%%%
\section{Conclusion}\label{sec:conclusion}
%%%%%%%%%%%%%%%%%%%%%%%%%%%%%%%%%%%%%%%%

The chiral SYK model serves as an interesting example of exactly solvable large-$N$ QFT in $1+1$ dimensions that do not possess Lorentz symmetry. Unlike the $0+1$-dimensional SYK model, it does not acquire conformal symmetry in the IR. However, as a consequence of the exact scaling symmetry, the system admits an exact solution of the two-point function and the interaction remains marginal when the SYK coupling strength is restricted to $0 \le J <2\pi$. In this work, we have studied two chiral SYK systems coupled by a relevant quadratic interaction that breaks the scaling and time-reversal symmetry of the system. The variant represents a tunnel junction between the two chiral SYK subsystems. The exact solution of the two-point function in chiral SYK model allows us to compute the correlators of the coupled SYK analytically when the two systems are weakly coupled. The DS equation is solved perturbatively to obtain the averaged thermal correlator $G^{\beta}_{AB}$ to leading order in $\mu$. The analytic expression involves a combination of complete elliptic integrals of the first kind. A consistent regularisation scheme corresponding to point splitting in the $x$ direction plays a crucial role in the computation involving real space integrals over the correlators. To validate our results, we compute the leading correction to the thermodynamic free energy of the coupled system using two independent methods. In contrast to the $0+1$-dimensional coupled SYK, the correlators do not decay exponentially with Euclidean time for large $\beta$, which implies that the coupled system does not acquire a mass gap at low temperature. Thus we find no evidence of a thermal phase transition occurring in the system. The leading correction to free energy is independent of the temperature; hence, the ground state entropy density remains same as that of $2N$ free Majorana fermions. However, the ground state of the coupled system lies below the ground state of the two decoupled chiral systems. The Euclidean two-point function is analytically continued to the retarded correlator and we find evidence of massless collective bosonic modes at zero temperature. This agrees with the general belief that the edge theory of a $2+1$ dimensional topological bulk is massless and aligns with the expected fermion pairing occurring in the bulk.

Chiral fermions arise in the edge theories of $2+1$-dimensional gapped topological phases of matter \cite{laughlin1981quantized,halperin1982quantized,haldane1988model,qi2006topological,moore1991nonabelions, read2000paired,qi2011topological,kitaev2006anyons}. In systems with $p$-wave fermion pairing that break parity and time-reversal symmetry, massless Majorana modes appear at the edge or attached to well-separated vortices \cite{read2000paired,qi2011topological}. The chiral SYK model offers a useful effective framework in which certain aspects of interacting Majorana edge dynamics can be explored near the Fermi level. A large number of chiral Majorana fermions also appear naturally in infinite layer quantum Hall systems \cite{balents1996chiral,naud2000fractional}. Within this perspective, the quadratic interaction studied here can be viewed as an interlayer tunnelling potential. Although the quadratic coupling appears nonlocal from the boundary viewpoint, similar structures can arise from local interactions in an underlying bulk description. Our analysis indicates that gapless chiral edge theories exhibit a form of robustness against relevant deformations that has no analog in $0+1$-dimensional SYK models. These observations may be useful for understanding tunnelling between Majorana edges with protected chirality and the role of bulk–boundary correspondence in topological condensed matter systems from a boundary perspective, although further work is needed to clarify these connections.

To determine the temperature regime in which our perturbative analytic treatment remains reliable and to extend our results beyond small values of $\mu$, it is important to perform a nonperturbative numerical analysis of the DS equations. Furthermore, one can study the overlap of the ground state of this coupled system with the thermofield double state and investigate its implication in the context of $2+1$ dimensional topological bulk. Although there is no indication of a thermal phase transition, a real-time formulation of the model is necessary to understand whether a quantum phase transition occurs here. This can also shed light on the lifetime of the quasiparticles and the existence of a plausible bulk dual. We leave these questions for future work.

%%%%%%%%%%%%%%%%%%%%%%%%%%%%%%%%%%%%%%%%
\section*{Acknowledgements}
%%%%%%%%%%%%%%%%%%%%%%%%%%%%%%%%%%%%%%%%
We would like to thank Pratik Das for useful discussions on various stages of this work. A. C. acknowledges financial support from the DST-INSPIRE Fellowship (IF200258), Department of Science and Technology, Government of India. The work of M. M. is supported by DST-FIST Grants No. SR/FST/LS-I/2020/621 and No. SR/FST/PSI-225/2016.

\appendix

\section{Regularising the free energy of chiral SYK system}\label{App:chiral_free_energy}

\begin{figure}[ht]
  \centering
  \begin{subfigure}[b]{0.45\textwidth}
    \centering
    \begin{overpic}[scale=0.065]{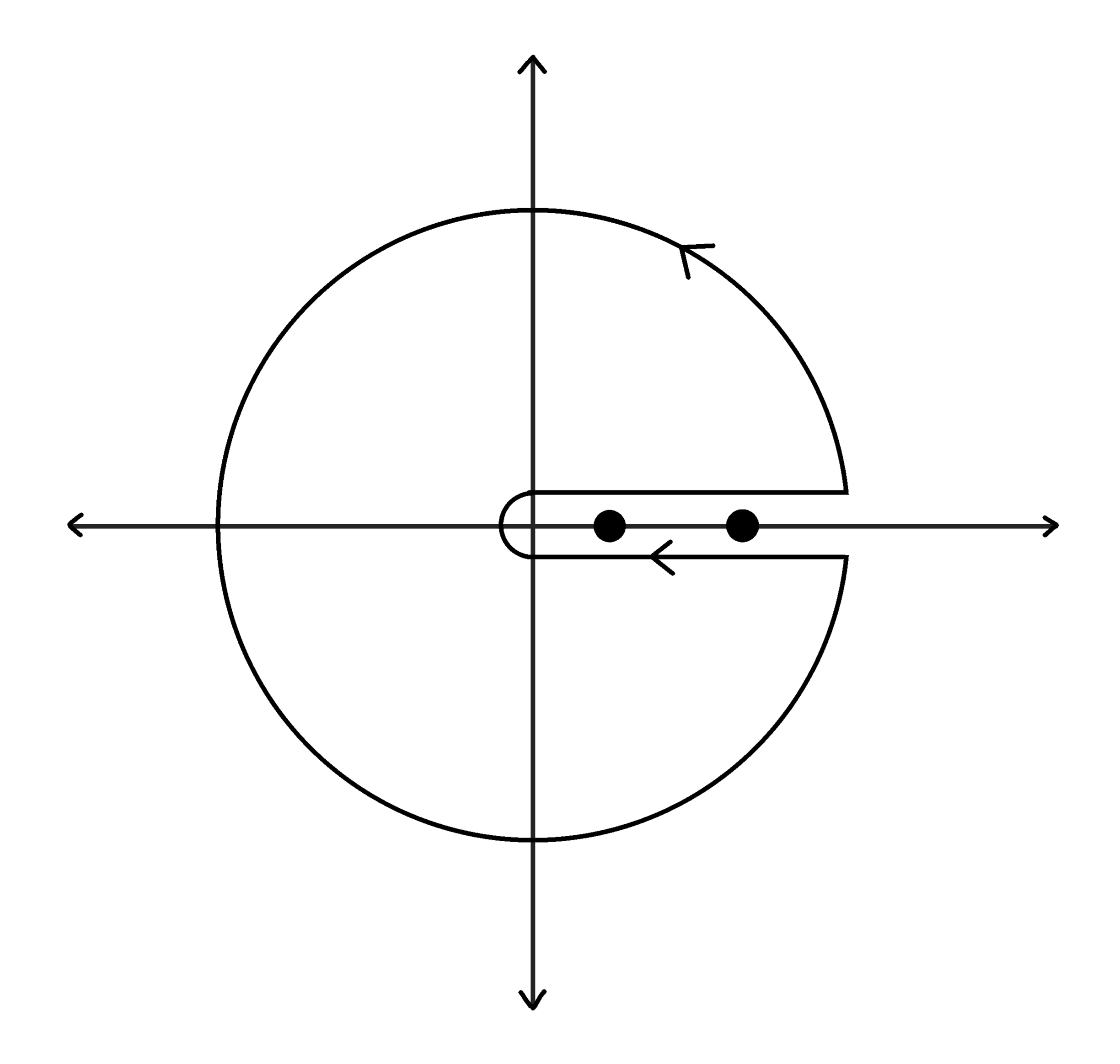} 
    \put(86,41){Re$(\tilde{z})$}
    \put(28,83){Im$(\tilde{z})$}
    \put(49,52){$e^{-u_+^{-1} x}$}
    \put(59,37){$e^{-u_-^{-1} x}$}
    \put(5,67){$|\tilde{z}|=1$}
\end{overpic}
    \caption*{(a)} 
  \end{subfigure}
  \hspace{0.7cm}
  \begin{subfigure}[b]{0.45\textwidth}
    \centering
    \begin{overpic}[scale=0.07]{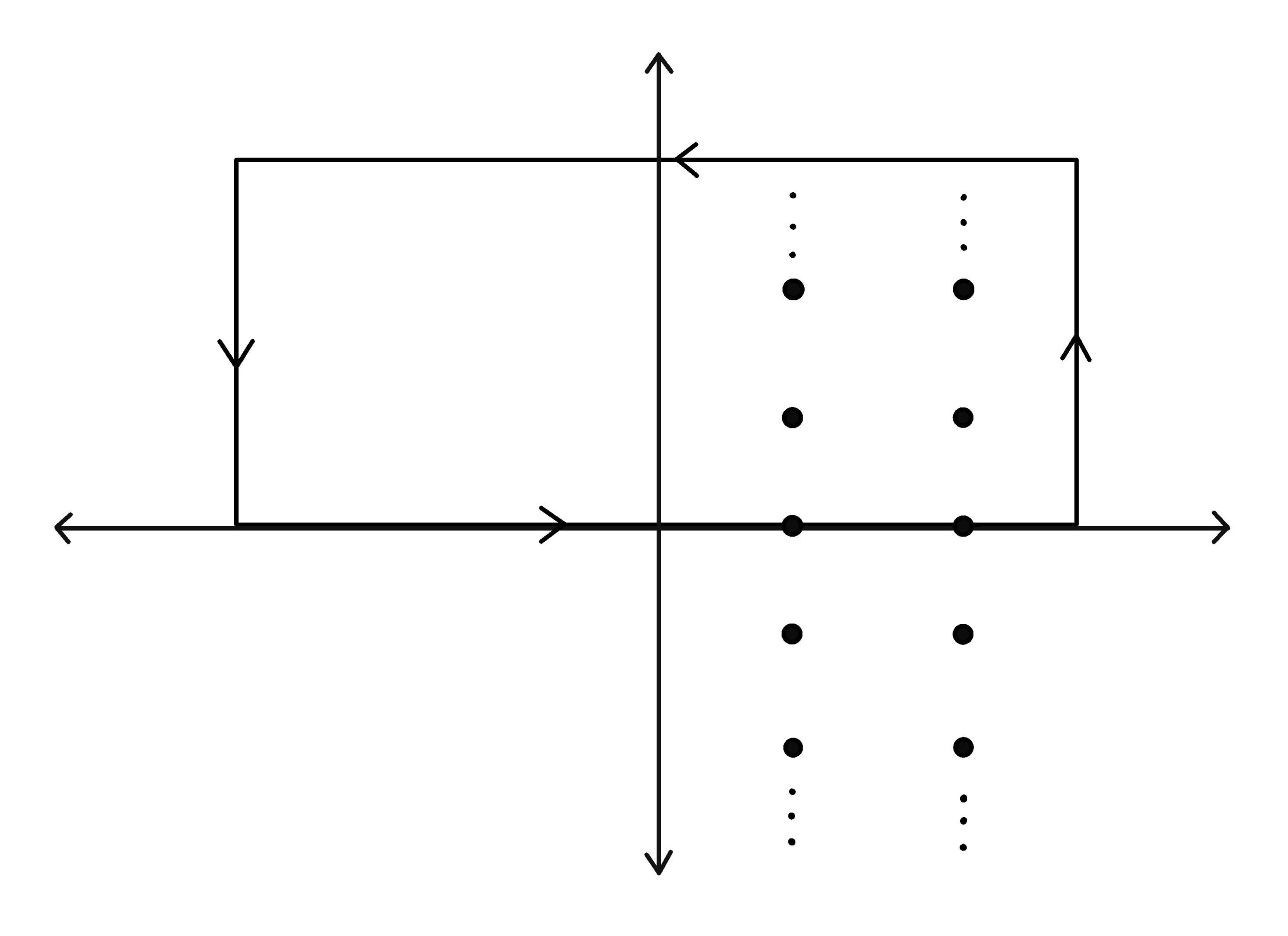} 
     \put(86,25){Re$(x)$}
     \put(36,64){Im$(x)$}
     \put(45,43){$u_- (t + 2in\pi)$}
     \put(67,35){$u_+ (t + 2in\pi)$}
     \put(15,60){\large$\tilde{\mathcal{C}}$}
     \end{overpic}
    \caption*{(b)}
  \end{subfigure}
  \caption{(a) The circular contour $|\tilde{z}|=1$ deformed to exclude the real $\tilde{z}$ axis where the poles are situated. (b) The square contour $\tilde{\mathcal{C}}$ which reduces to the $x$ integral when the length of the sides reach infinity. The two poles shown on the real $x$ axis lie infinitesimally above for Im$(t) = \epsilon$ and infinitesimally below for Im$(t) = -\epsilon$.}
  \label{chiral_free_en}
\end{figure}

The thermodynamic free energy $F_{\mr{chiral}}$ and entropy density $\SSS_{\mr{chiral}}$ of the chiral SYK model in the large-$N$ limit is obtained by putting saddle-point solutions of the bilinears $G$ and $\Sigma$ into the disorder-averaged partition function. For a chiral SYK system with size $L$, the thermal quantities can be expressed as
\beq
 -\frac{\beta F_{\mr{chiral}}}{L} = \frac{\log \langle Z_{\mr{chiral}} \rangle}{L} \quad \mr{and} \quad \SSS_{\mr{chiral}} = (1-\beta \partial_{\beta}) \frac{\log \langle Z_{\mr{chiral}} \rangle}{L}.
\eeq
To avoid dealing with the $\tr \log$ term in $\log \langle Z_{\mr{chiral}} \rangle$ [Eq. \eqref{eq:chiral_logZ}], it is convenient to compute $\partial_J \log \langle Z_{\mr{chiral}} \rangle $ and integrate over $J$. The terms proportional to $\partial_J G$ and $\partial_J \Sigma $ vanish at the saddle point, which leads to
\beq
\label{eq:partial_JJ}
 \frac{\partial_{J} \log \langle Z_\mr{chiral} \rangle}{L} = \frac{NJ}{64\beta^3 u_+^2 u_-^2} \int_{-\beta/2}^{\beta/2} d\tau \int_{-\infty}^{\infty} dx \; \dfrac{1}{\sin^2 \left[ \frac{\pi}{\beta}(\tau-iu_+^{-1}x) \right]\sin^2 \left[ \frac{\pi}{\beta}(\tau-iu_-^{-1}x) \right]}.
\eeq
Here we use the same method used to perform the spacetime integrals at finite temperature throughout the paper. Rescaling the coordinates by a factor of $2\pi/\beta$ and defining a complex variable $\tilde{z}=e^{i\tau}$, the $\tau$ integral reduces to a integral along the unit circle $|\tilde{z}| = 1$ excluding the segment from $\tilde{z}=1-i\epsilon$ to $1+i\epsilon$, which is consistent with the real-space UV cutoff $|\tau| > \epsilon $. The contour is closed by deforming the circle around the real $\tilde{z}$ axis where the poles lie as shown in Fig. \ref{chiral_free_en}.
Since the integrand is analytic inside the contour, the integral along the unit circle can be expressed in terms of the integrals along the lines infinitesimally above and below the real $\tilde{z}$ axis. We represent the integral above and below the real $\tilde{z}$ axis as $\tilde{I}_1$ from $t=\infty -i\epsilon$ to $t=-i\epsilon$ and $\tilde{I}_2$ from $t=i\epsilon$ to $t=\infty + i\epsilon$, respectively. Introducing the Minkowski-time $t$ as $\tilde{z}= e^{-t}$, we can reduce Eq. \eqref{eq:partial_JJ} to
\beq
 \frac{\partial_{J} \log \langle Z_\mr{chiral} \rangle}{L} = -\tilde{I}_1 -\tilde{I}_2 = \frac{iNJ}{256\beta \pi^2 u_+^2 u_-^2} \left( \int_{-i\epsilon}^{\infty-i\epsilon}dt + \int_{\infty + i\epsilon}^{i\epsilon}dt \right) \tilde{I}_x ,
\eeq
where the $x$ integral,
\beq
 \tilde{I}_x = \int_{-\infty}^{\infty} \frac{dx}{\sinh^2 \left( \frac{t - u_+^{-1}x}{2} \right) \sinh^2 \left( \frac{t - u_-^{-1}x}{2} \right) }.
\eeq
The integral can also be represented by a contour integral of the form
\beq
 \tilde{I}_{\tilde{\mathcal{C}}} = \int_{\tilde{\mathcal{C}}} \frac{dx \, e^{-\mr{Im}(x)}}{\sinh^2 \left( \frac{t - u_+^{-1}x}{2} \right) \sinh^2 \left( \frac{t - u_-^{-1}x}{2} \right) },
\eeq
where $\tilde{\mathcal{C}}$ is a square contour as shown in Fig. \ref{chiral_free_en}. It is straightforward to see that the integrand vanishes on the lines $\mr{Re}(x) \rightarrow \pm \infty$ and $\mr{Im}(x) \rightarrow - \infty$ . For fixed $t$, the second-order poles are situated at $x=u_\pm (t + 2n\pi i)$ where $n \in \mathbb{Z}$. Since the cutoff parameter puts a small imaginary part in $t$, the poles for $n \geq 0$ lie inside the contour for $\tilde{I}_2$, whereas only the $n>0$ poles lie inside $\tilde{\mathcal{C}}$ for $\tilde{I_1}$. All pole contributions except $n=0$ cancel out due to the opposite limits of $t$ in $\tilde{I}_1$ and $\tilde{I}_2$. The surviving contribution gives
\beq
 \frac{\partial_{J} \log \langle Z_\mr{chiral} \rangle}{L} = \frac{iNJ}{256\beta \pi^2 u_+^2 u_-^2} \int_{\infty + i\epsilon}^{i\epsilon}dt 2\pi i \left[ \mr{Res}(x=u_+ t) + \mr{Res}(x=u_- t) \right],
\eeq
where
\beal
 \mr{Res}(x=u_+ t) &= \frac{4u_+^2}{u_-} \coth \left[ \frac{(u_- -u_+)t}{2u_-} \right] \mr{cosech}^2 \left[ \frac{(u_- -u_+)t}{2u_-} \right], \\
 \mr{Res}(x=u_- t) &= \frac{4u_-^2}{u_+} \coth \left[ \frac{(u_+ -u_-)t}{2u_+} \right] \mr{cosech}^2 \left[ \frac{(u_+ -u_-)t}{2u_+} \right].
\eeal
Finally, integrating over $t$ yields
\beq
 \frac{\partial_{J} \log \langle Z_\mr{chiral} \rangle}{L} = -\frac{NJ}{32 \beta \pi (u_+ - u_-)} \left[ \frac{1}{u_-^2} \mr{cosech}^2 \left( \frac{(u_+ - u_-)i\epsilon}{u_-} \right) - \frac{1}{u_+^2} \mr{cosech}^2 \left( \frac{(u_+ - u_-)i\epsilon}{u_+} \right) \right].
\eeq
The $\epsilon^{-2}$ divergent piece is removed from the Taylor expansion. The cutoff independent finite piece gives the regularised value
\beq
 \frac{\partial_{J} \log \langle Z_\mr{chiral} \rangle}{L} = \frac{N}{96}\left( u_-^{-2} - u_+^{-2} \right) \beta^{-1}.
\eeq
This leads to the temperature-dependent piece in free energy density for the chiral SYK model (Eq. \eqref{eq:JJ_int}). The free energy density is obtained as
\beq
 \frac{F_{\mr{chiral}}}{L} = a_0(J,\epsilon^{-2}) + \frac{N\pi}{48 \beta^2}(u_-^{-1} + u_+^{-1} - 2) + \OO(\epsilon^2,\beta^{-4}),
\eeq
where $a_0$ represents the divergent piece.

\section{Contours in computing the two-point function}
\label{App:two-point_fn}
 In this Appendix, we present explicit computation of the leading correction to the two-point function $G_{AB}^{\beta}(\tau,x)$ given by
 \beq
   G_{AB}^{\beta}(\tau,x) = i\mu \int_{0}^{\beta}d\tau_1 \int_{-\infty}^{\infty}dx_1\  G_{AA}^{\beta}(\tau-\tau_1,x-x_1)G_{BB}^{\beta}(\tau_1,x_1),
  \eeq
   The real space integral needs to be regularised by imposing the UV cutoff $|\tau_1| \geq \epsilon$ and $|\tau-\tau_1| \geq \delta$ consistent with the point splitting in the $x$ direction as discussed in Sec. \ref{subsec:largeNprop}. Thus, the $\tau_1$ integral with the regularisation parameters $\epsilon$ and $\delta$ needs to be rewritten as
 \beq
   \int_0^\beta d\tau_1 \rightarrow \int_{\epsilon}^{\tau - \delta} d\tau_1 + \int_{\tau + \delta}^{\beta - \epsilon} d\tau_1 .
  \eeq
  
  The integrand in \eqref{eq:GLRcorrection} has infinitely many branch points; hence, the integral is very difficult to deal with. We adopt a method analogous to that used in Ref. \cite{Lian_2019}, which allows us to reduce the problem to integrals around a finite number of branch cuts. We first transform the integration over $\tau_1$ in \eqref{eq:GLRcorrection} to an integration along the counterclockwise circular contour of unit radius by rescaling the coordinates by a factor of $2\pi/\beta$ and substituting $z_1 = e^{i\tau_1}$. The UV cutoff in $\tau_1$ translates to $\epsilon \le \mr{Arg}(z_1) \le \tau - \delta$ and $\tau + \delta \le \mr{Arg}(z_1) \le \beta - \epsilon$ on the circular contour. For convenience, we shall absorb any multiplicative constants into the cutoff parameters $\epsilon$ and $\delta$ without explicitly stating so. Also, we keep only the linear order terms in $\epsilon$ and $\delta$ and express the correlator as
  \beal
   G_{AB}^{\beta}(\tau,x) = \mu P(\tau,x) \int_{|z_1|=1} \frac{dz_1}{z_1} \int_{-\infty}^{\infty} dx_1 \Bigg[ \left(zz_1^{-1} e^{u_+^{-1}(x-x_1)} - 1 \right)& \left(zz_1^{-1} e^{u_-^{-1}(x-x_1)}-1 \right) \\
   \left(z_1 e^{u_+^{-1}x_1}-1 \right)& \left(z_1 e^{u_-^{-1}x_1}-1 \right) \Bigg]^{-1/2},
  \eeal
  where we have used $z=e^{i\tau}$ and $P(\tau,x) = -\dfrac{\exp\left[\frac{i\tau}{2} + (u_+^{-1}+u_-^{-1})\frac{x}{4} \right]}{4\pi^2 u_+ u_-}$.

\begin{figure}[ht]
   \centering
   \includegraphics[scale=0.55]{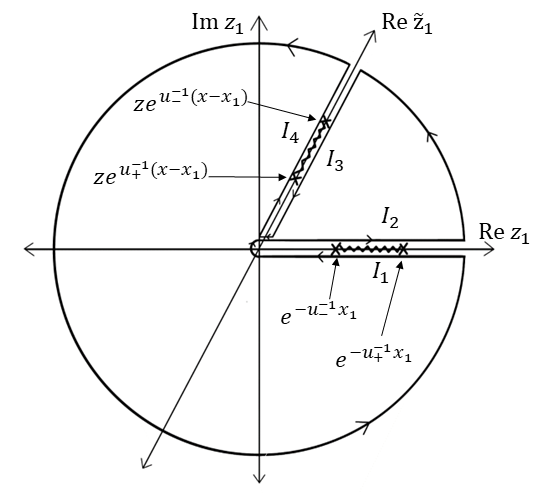}
   \caption{The unit circle contour is deformed to avoid the poles and the branch cuts. With no poles or cuts inside, the contour reduces to the four integral, $-I_1-I_2-I_3-I_4$.}
   \label{deformedcont}
  \end{figure}

  The integrand has a simple pole at $z_1=0$ and branch points at $\displaystyle{z_1 = ze^{u_{\pm}^{-1}(x-x_1)},e^{-u_{\pm}^{-1}x_1}}$ for fixed $x$. We can deform the contour $|z_1|=1$ into a contour closely bypassing the positive real $z_1$ axis, the $z_1 = 0$ point and the $z_1 = r z = r e^{i\tau},\ (0<r\le 1)$ line as shown in Fig. \ref{deformedcont}. The pole and all the branch points lie outside of the deformed contour.
  Then the unit circle contour will be reduced to the four integrals, $G_{AB}^{\beta} = -I_1 -I_2 -I_3 -I_4$.

  We denote $z= e^{-t}$ and $z_1 = e^{-t_1}$ while computing $I_1$ and $I_2$, where $t$ and $t_1$ denote the Minkowski time variables obtained from the Wick rotation of the Euclidean times, $\tau = it$ and $\tau_1 = it_1$. The deformed parts of the $z_1$ contour integral from $e^{-i\epsilon}$ to $0$ below the real axis, denoted by $I_1$, and from $0$ to $e^{i\epsilon}$ above the real axis, denoted by $I_2$, are then equivalent to integrals of $t_1$ from $i\epsilon$ to $\infty +i\epsilon$ and from $\infty -i\epsilon$ back to $-i\epsilon$, respectively, where $\epsilon \to 0^+$. It is easy to see that the imaginary part of $t_1$ is infinitesimally positive for $I_1$ and is infinitesimally negative for $I_2$.

  While computing $I_3$ and $I_4$, we denote $z_1 = z e^{-\bar{t}_1}$, where $\bar{t}_1$ is obtained by Wick rotating $(\tau_1 - \tau)$, $\bar{t}_1 = -i(\tau_1 -\tau)$. The deformed parts of the $z_1$ contour integral from $z e^{-i\delta}$ to $0$, denoted by $I_3$ (close to $I_2$ line but not intersecting), and from $0$ to $z e^{i\delta}$, denoted by $I_4$, are equivalent to integrals of $\bar{t}_1$ from $i\delta$ to $\infty +i\delta$ and from $\infty -i\delta$ back to $-i\delta$, respectively, where $\delta \to 0^+$. Again, it is straightforward to argue that the imaginary part of $\bar{t}_1$ is infinitesimally positive for $I_3$ and is infinitesimally negative for $I_4$. The integral can be rewritten as integrals with respect to $t_1$ and $\bar{t}_1$,
  \beq
   \label{deformedt1}
   G_{AB}^{\beta} = \mu P \left[ \left(\int_{i\epsilon}^{\infty +i\epsilon}dt_1 + \int_{\infty -i\epsilon}^{-i\epsilon}dt_1\right) I_a + \left(\int_{i\delta}^{\infty +i\delta}d\bar{t}_1 + \int_{\infty -i\delta}^{-i\delta}d\bar{t}_1\right) I_b \right],
  \eeq
  where the integral with respect to $x_1$ is expressed with the variable $t_1$ for $I_1$ and $I_2$,
  \beq
   \label{firstx1integral}
   I_a = \int_{-\infty}^{\infty} \frac{dx_1}{\sqrt{\left( e^{t_1-t+u_+^{-1}(x-x_1)}-1 \right)\left( e^{t_1-t+u_-^{-1}(x-x_1)}-1 \right) \left( e^{-t_1+u_+^{-1}x_1}-1 \right) \left( e^{-t_1+u_-^{-1}x_1}-1 \right) }},
  \eeq
  and it is expressed with the variable $\bar{t}_1$ for $I_3$ and $I_4$,
  \beq
   \label{secondx1integral}
   I_b = \int_{-\infty}^{\infty} \frac{dx_1}{\sqrt{\left( e^{-\bar{t}_1-t+u_+^{-1}x_1}-1 \right)\left( e^{-\bar{t}_1-t+u_-^{-1}x_1}-1 \right) \left( e^{\bar{t}_1+u_+^{-1}(x-x_1)}-1 \right) \left( e^{\bar{t}_1+u_-^{-1}(x-x_1)}-1 \right)}}.
  \eeq

\begin{figure}[ht]
  \centering
  \includegraphics[scale=0.6]{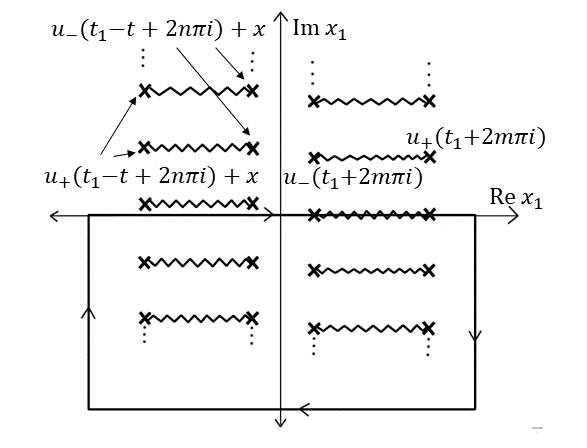}
  \caption{Countably infinite branch cuts of the integrand of $I_a$. Depending on the imaginary part of $t_1$, the cut shown on the Re $x_1$ axis will lie inside or outside of the rectangular contour $C$.}
  \label{infbranchcuts}
 \end{figure}

  The idea behind reducing \eqref{eq:GLRcorrection} to \eqref{deformedt1} is to argue the cancellation of infinitely many branch cut contributions due to the opposite limits in the $t_1$ integrals. But to write $I_a$ and $I_b$ as integrals around the branch cuts, we need to find a suitable contour. We consider a rectangular contour $C$ as shown in Fig. \ref{infbranchcuts} and define the contour integral,
 \beq
  \label{squarecontour}
  I_C = \int_{C} \frac{dx_1 e^{\mathrm{Im}(x_1)}}{\sqrt{\left( e^{t_1-t+u_+^{-1}(x-x_1)}-1 \right)\left( e^{t_1-t+u_-^{-1}(x-x_1)}-1 \right) \left( e^{u_+^{-1}x_1 -t_1}-1 \right)\left( e^{u_-^{-1}x_1-t_1}-1 \right)}}.
 \eeq
 Among the four sides of the contour $C$, the side on the real $x_1$ axis reduces to $I_a$ since $e^{\mr{Im}(x_1)} \rightarrow 1$. The contributions from the other three sides vanish when the length of the sides are taken to infinity, and thus $I_C$ reduces to $I_a$.
 
 The integrand in \eqref{squarecontour} has infinitely many countable branch cuts along the line segments from $u_-(t_1+2m\pi i)$ to $u_+(t_1+2m\pi i)$ and from $u_-(t_1-t+2n\pi i) + x$ to $u_+(t_1-t+2n\pi i) + x$ for a fixed $t$ as represented in Fig. \ref{infbranchcuts}. Since we have imposed the UV cutoff on the Euclidean times, $|\tau_1|>\epsilon$ and $|\tau| > \epsilon'$, the imaginary part of the Minkowski time variable is nonzero on the integration lines. Assuming $\epsilon'>2\epsilon$, it is straightforward to check that all the branch cut contributions except the cut corresponding to $m=0$ cancel out due to opposite limits of $t_1$ in the integrals $I_1$ and $I_2$. The $m=0$ branch cut from $u_- t_1$ to $u_+ t_1$ lies inside the contour when Im$(t_1)=-\epsilon$ (for $I_2$) and lies outside when Im$(t_1)=\epsilon$ (for $I_1$). The surviving contribution to $-I_1 -I_2$ is obtained by reducing $I_C$ into a contour closely surrounding the $m=0$ cut with Im$(t_1)=-\epsilon$. We restrict ourselves to one thermal circle $0< \tau (\beta/2\pi)<\beta$ for a consistent choice of branch for the square roots in integrals $I_a$ and $I_b$. In this range of $\tau$, the other branch cut from $u_-(t_1-t) + x$ to $u_+(t_1-t) + x$ lies below the contour encircling the surviving branch. Choosing the branch of the square roots accordingly, the first two integrals in Eq. \eqref{deformedt1} reduce to
 \beal
  \label{firstint}
  -I_1 -I_2 = 2\mu P \int_{-i\epsilon}^{\infty -i\epsilon} dt_1 \int_{u_- t_1}^{u_+ t_1}dy \Bigg[ \left( e^{t_1 -t+u_+^{-1}(x-y)}-1 \right)& \left( 1-e^{t_1 -t+u_-^{-1}(x-y)} \right) \\
  \left( 1-e^{-t_1 +u_+^{-1}y} \right)& \left( e^{-t_1 +u_-^{-1}y}-1 \right)\Bigg]^{-\frac{1}{2}},
 \eeal
 where $y = \mathrm{Re}(x_1)$. The expression can be simplified further by defining the coordinates $y_+= t_1 -u_+^{-1}y$ and $y_- = u_-^{-1}y-t_1$. The double integral in \eqref{firstint} decouples into
 \beal
  \label{secondint}
  -I_1-I_2 = \frac{2\mu u_+ u_- P}{u_+ - u_-} \int_{-i\epsilon}^{\infty -i\epsilon}dy_+ &\left[ \left( c_+^2 e^{y_+}-1 \right)\left( 1-e^{-y_+} \right) \right]^{-\frac{1}{2}} \\
  &\int_{i\epsilon}^{\infty +i\epsilon}dy_- \left[ \left( 1-c_-^2 e^{-y_-} \right)\left( e^{y_-}-1 \right) \right]^{-\frac{1}{2}}.
 \eeal
 where $c_{\pm}(\tau,x) = \exp\left[(u_{\pm}^{-1}x + i\tau)/2 \right] $. Now we can reduce the integrals to the standard form of complete elliptic integral of the first kind by substituting $s_{\pm} = e^{-y_\pm /2}$. The regularisation parameter $\epsilon$ keeps track of the branches of square roots in the integrals. Eq. \eqref{secondint} is rewritten as
 \beq
 \label{eq:finalint}
  -I_1-I_2 = \frac{8\mu u_+ u_- P}{u_+ - u_-} \int_0^{1+i\epsilon} \frac{ds_+}{s_+ \sqrt{\left(c^2_+ s_+^{-2} -1\right)\left(1-s_+^2 \right)}} \int_0^{1-i\epsilon} \frac{ds_-}{s_- \sqrt{\left(1-c_-^{2}s_-^2 \right)\left(s_-^{-2}-1 \right)}},
 \eeq
 In the limit $\epsilon \to 0$, the resulting integrals can be expressed as,
 \beq
  \label{eq:reducedint1}
  -I_1-I_2 = \frac{\mu \tilde{P}}{c_+}\ K\left( \dfrac{1}{c_+} \right)K\left( c_- \right),
 \eeq
 where
 \beq
  \tilde{P}(\tau,x) = -\frac{2}{\pi^2 (u_+ - u_-)} \exp \left[ \frac{i\tau}{2} + \frac{1}{4}\left(u_+^{-1}+u_-^{-1}\right)x \right]
 \eeq
 and $K$ denotes the complete elliptic integral of first kind \cite{NIST:DLMF,lawden2013elliptic},
 \beq
  \label{eq:elliptic_integral}
  K(c) = \int_0^1 ds \left[ (1-s^2)(1-c^2 s^2) \right]^{-1/2}.
 \eeq

 Now, we focus on to the other branch cut contribution $-I_3-I_4$ to the two-point function $G^{\beta}_{AB}$. Following the technique used previously, we define a rectangular contour closing in the lower half-plane which reduces to $I_b$ if the length of each side reaches infinity. The imposed UV cutoff translates to a small imaginary part in $\bar{t}_1$ which is positive for $I_3$ and negative for $I_4$. Only the contribution of the branch cut from $x+u_-\bar{t}_1$ to $x+u_+\bar{t}_1$ remains after cancellation of all other branch cut contributions of $I_b$ in $I_3$ and $I_4$ due to opposite limits of $\bar{t}_1$ in integrals $I_3$ and $I_4$. Contrary to the previous case, the other branch cut inside the thermal circle $0<\tau(\beta/2\pi)<\beta$ from $u_-(t+\bar{t}_1)$ to $u_+(t+\bar{t}_1)$ lies above the contour encircling the surviving branch. A consistent branch choice of the square root functions in $I_b$ leads to
 \beal
  -I_3 -I_4 = -2\mu P \int_{-i\delta}^{\infty -i\delta} d\bar{t}_1 \int_{x+u_- \bar{t}_1}^{x+u_+ \bar{t}_1}dy \Bigg[ \left( e^{\bar{t}_1 +u_+^{-1}(x-y)}-1 \right)& \left( 1-e^{\bar{t}_1 +u_-^{-1}(x-y)} \right) \\
  \left( 1-e^{-\bar{t}_1-t +u_+^{-1}y} \right)& \left( e^{-\bar{t}_1 -t+u_-^{-1}y}-1 \right)\Bigg]^{-\frac{1}{2}}.
 \eeal
 Again, the resulting double integral can be decoupled by defining $\bar{y}_+= \bar{t}_1 + u_+^{-1}(x-y)$ and $\bar{y}_- = -\bar{t}_1 - u_-(x-y)$. The substitution $r_\pm = e^{-\bar{y}_\pm /2}$ reduces it to a standard form,
 \beq
  -I_3-I_4 = -\frac{8\mu u_+ u_- P}{u_+ - u_-} \int_0^{1+i\delta} \frac{dr_+}{r_+ \sqrt{\left(1-c^2_+ r_+^{2} \right)\left(r_+^{-2} -1 \right)}} \int_0^{1-i\delta} \frac{dr_-}{r_- \sqrt{\left(c_-^{2}r_-^{-2} -1\right)\left(1- r_-^{2} \right)}},
 \eeq
 where each of the integrals can be represented by complete elliptic integrals of the first kind after taking the limit $\delta \to 0$. The second contribution to $G^{\beta}_{AB}$ becomes
 \beq
 \label{eq:reducedint2}
  -I_3-I_4 = -\frac{\mu \tilde{P}}{c_-}\ K\left( \dfrac{1}{c_-} \right)K\left( c_+ \right).
 \eeq
 Finally, restoring the coordinates by rescaling with $2\pi/\beta$ and combining \eqref{eq:reducedint2} with Eq. \eqref{eq:reducedint1} we get the resulting $G_{AB}^{\beta}$ to the linear order in $\mu$ for nonzero imaginary-time separation,
 \beq
  G_{AB}^{\beta}(\tau,x) = \frac{2\mu}{\pi^2 (u_+ -u_-)}  \left[ \sqrt{\frac{c_+}{c_-}} K\left( \frac{1}{c_-} \right)K\left( c_+ \right) - \sqrt{\frac{c_-}{c_+}} K\left( \frac{1}{c_+} \right)K\left(c_- \right)\right],
 \eeq
 where 
 \beq
  c_\pm = \exp \left[ \frac{\pi}{\beta}(u_\pm^{-1}x +i\tau) \right].
 \eeq

\section{Contours in computing the leading correction to free energy}
\label{App:free_en_correction}
 In this Appendix, we provide the detailed derivation of $G_{AB}^{\beta}(0,0)$ using Eq. \eqref{eq:GLRcorrection}, which yields the leading-order correction to the free energy of the coupled model. The thermal two-point function $G_{AB}^{\beta}$ as expressed in Eq. \eqref{eq:G_AB_zero} reads
 \beq
  G_{AB}^{\beta}(0,0) = -\frac{i\mu}{4\beta^2 u_+u_-} \int_{\epsilon_\tau}^{\beta -\epsilon_\tau}d\tau \int_{-\infty}^{\infty} \frac{dx}{\sin \left[ \frac{\pi}{\beta}(\tau-iu_+^{-1}x) \right] \sin \left[ \frac{\pi}{\beta}(\tau-iu_-^{-1}x) \right]},
 \eeq
 where $\epsilon$ is the real space UV cutoff, $|\tau|>\epsilon$. The integrand has infinite number of poles and can be tackled using the same method used in Appendix \ref{App:chiral_free_energy}.

 \begin{figure}[ht]
  \centering
  \begin{subfigure}[b]{0.45\textwidth}
    \centering
    \begin{overpic}[scale=0.065]{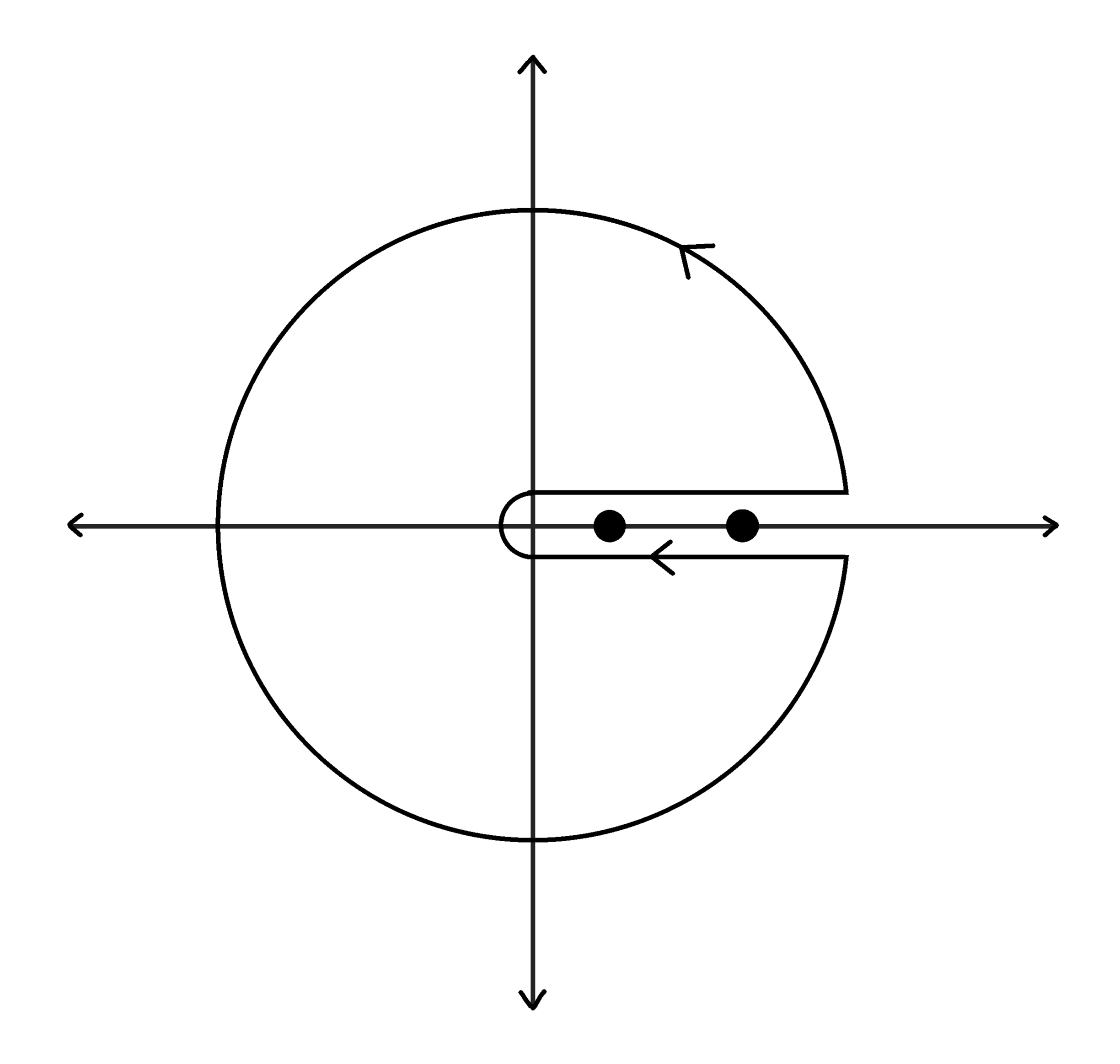} 
    \put(86,41){Re$(z')$}
    \put(28,83){Im$(z')$}
    \put(49,52){$e^{-u_+^{-1} x}$}
    \put(59,37){$e^{-u_-^{-1} x}$}
    \put(5,67){$|z'|=1$}
\end{overpic}
    \caption*{(a)} 
  \end{subfigure}
  \hspace{0.7cm}
  \begin{subfigure}[b]{0.45\textwidth}
    \centering
    \begin{overpic}[scale=0.07]{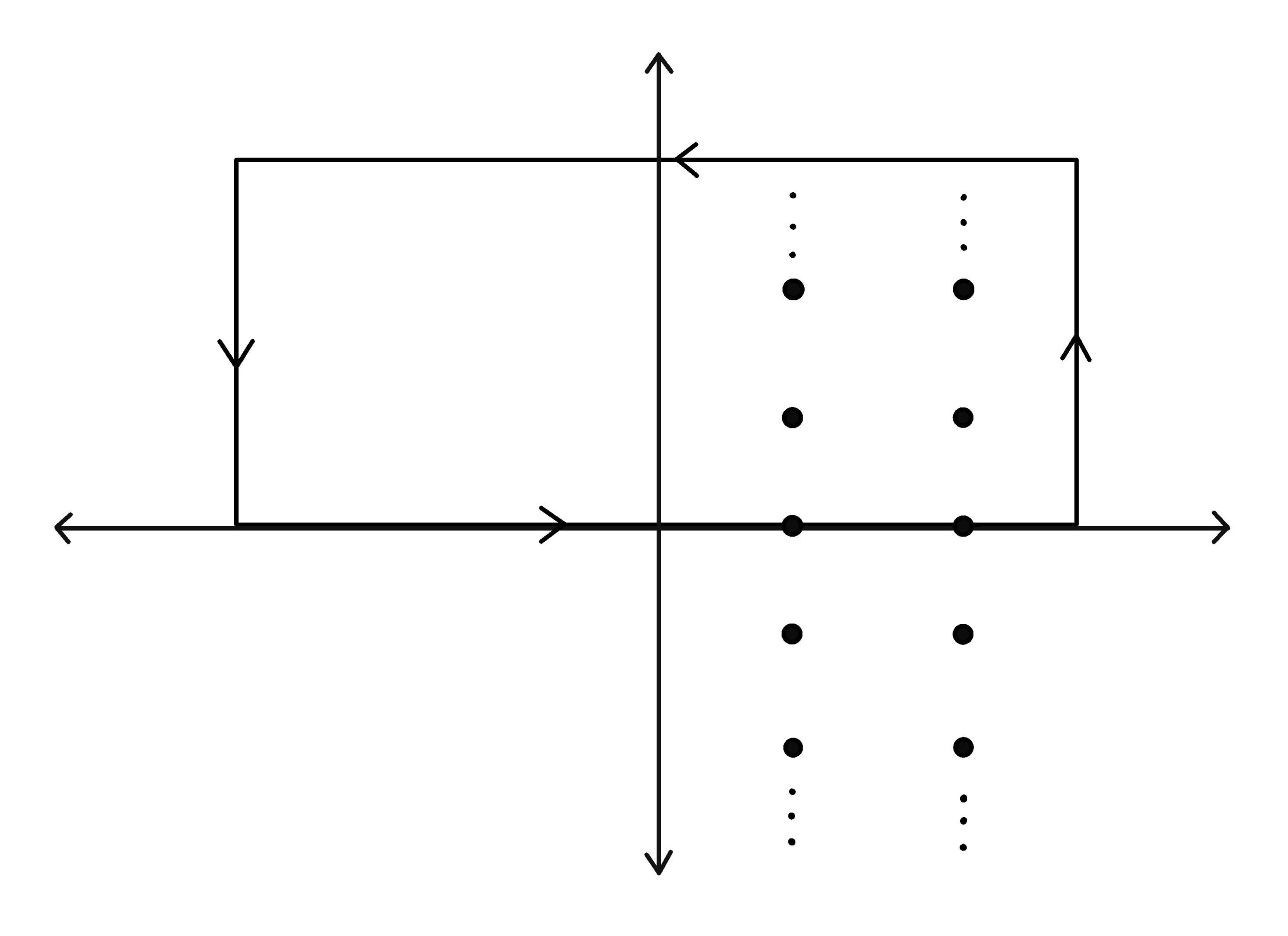} 
     \put(86,25){Re$(x)$}
     \put(36,64){Im$(x)$}
     \put(45,43){$u_- (t + 2in\pi)$}
     \put(67,35){$u_+ (t + 2in\pi)$}
     \put(15,60){\large$\mathcal{C}'$}
     \end{overpic}
    \caption*{(b)}
  \end{subfigure}
  \caption{(a) The circular contour $|z'|=1$ deformed to exclude the poles situated on real $\tilde{z}$ axis. (b) The infinite square contour $\mathcal{C}'$ reduces to the $x$ integral. The two poles shown on the real $x$ axis lie infinitesimally above for Im$(t) = \epsilon$ and infinitesimally below for Im$(t) = -\epsilon$.}
  \label{contours_free_en}
\end{figure}
  For fixed $x$, the poles are situated at $\tau = iu_\pm x$. Rescaling the coordinates by $(\tau,x) \rightarrow \frac{\beta}{2\pi}(\tau,x)$, the integral transforms into a contour integral in the complex $z'=e^{i\tau}$ plane along the unit circle $|z'| = 1$ with the imposed UV cutoff $\epsilon \le \mr{Arg}(z') \le 2\pi -\epsilon$, and the poles are situated on the real $z'$ axis. The contour is deformed to keep the real $z'$ axis outside the region closed by the contour (see Fig. \ref{contours_free_en}). Since the integrand is analytic inside the contour, the integral in Eq. \eqref{eq:G_AB_zero} can be expressed using the line integrals surrounding the real $z$ axis. Introducing the Minkowski-time variable $t=-i\tau$, we can write
  \beq
  \label{eq:t_integral}
   G_{AB}^{\beta}(0,0) = -\frac{\mu}{16 \pi^2 u_+ u_-} \left( \int_{\infty -i\epsilon}^{-i\epsilon} dt + \int_{i\epsilon}^{\infty +i\epsilon} dt \right) I_{x} \; ,
  \eeq
  where the $x$ integral is represented by
  \beq
   I_x = \int_{-\infty}^{\infty} dx \left[ \sinh \left( \frac{t -u_+^{-1}x}{2} \right) \sinh \left( \frac{t -u_-^{-1}x}{2} \right) \right]^{-1}.
  \eeq
  To compute $I_x$, we define a rectangular contour $\mathcal{C'}$ with the integral
  \beq
   I_{\mathcal{C'}} = \int_{\mathcal{C'}} dx e^{-\mr{Im}(x)} \left[ \sinh \left( \frac{t -u_+^{-1}x}{2} \right) \sinh \left( \frac{t -u_-^{-1}x}{2} \right) \right]^{-1}
  \eeq
  which reduces to $I_x$ in the limit where the sides of the rectangle reach infinity. The integrand of $I_{\mathcal{C'}}$ has infinitely many simple poles situated at $x=u_\pm (t + 2 i n \pi), $ where $n$ is an integer. All the pole contributions except $n=0$ cancel due to the opposite limits of $t$ in \eqref{eq:t_integral}. The poles corresponding to $n=0$, $x=u_\pm t$, lie inside the contour when $\mr{Im}(t) = \epsilon$, but lie outside when $\mr{Im}(t) = -\epsilon$. Considering only the surviving residues of the poles $x=u_\pm t$ for $\mr{Im}(t) = \epsilon$, we arrive at
  \beal
   G_{AB}^{\beta}(0,0) &= -\frac{\mu}{16 \pi^2 u_+ u_-} \int_{i\epsilon}^{\infty +i\epsilon} dt \; 2\pi i \left[ \mr{Res} \left(I_{\mathcal{C'}},x=u_+ t \right) + \mr{Res} \left( I_{\mathcal{C'}},x=u_- t \right) \right] \\
   &= -\frac{i\mu}{8 \pi u_+ u_-} \int_{i\epsilon}^{\infty +i\epsilon} dt \; \left[ 2u_+ \mr{cosech} \left\lbrace \frac{(u_+ -u_-)t}{2u_-} \right\rbrace - 2u_- \mr{cosech}\left\lbrace \frac{(u_+ -u_-)t}{2u_+} \right\rbrace \right] \\
   &= -\frac{i\mu}{2 \pi (u_+ -u_-)}  \left[ \log \tanh \left\lbrace \frac{(u_+ -u_-)t}{4u_-} \right\rbrace - \log \tanh \left\lbrace \frac{(u_+ -u_-)t}{4u_+} \right\rbrace\right]_{t=i\epsilon}^{t \rightarrow \infty + i\epsilon}.
  \eeal
  The upper limit is well defined and easily obtained to be zero. The infinitesimal parameter $\epsilon$ plays a crucial role while taking the lower limit smoothly. Ignoring the higher powers of $\epsilon$, we find the regularised value of the two-point function at the coincident point as
  \beq
   G_{AB}^{\beta}(0,0) = -\frac{i\mu}{2 \pi (u_+ -u_-)} \log \left( \frac{u_-}{u_+} \right).
  \eeq
  Finally, using this regularised value and the antisymmetry condition $G_{BA}^{\beta}(0,0)=-G_{AB}^{\beta}(0,0)$, we obtain the leading correction to the thermal free energy from Eq. \eqref{eq:free_ener_corr1},
  \beq
   \frac{\Delta F}{L} = -\frac{\mu^2 N}{2 J} \log \left( \frac{u_+}{u_-} \right).
  \eeq

%%%%%%%%%%%%%%%%%%%%%%%%%%
\bibliographystyle{utphys}
\bibliography{testbib}
\end{document}